\documentclass[12pt]{article}
\input epsf
\usepackage{psfrag}
\usepackage{graphicx}
\catcode`\@=11
\def\@citex[#1]#2{%
\if@filesw \immediate \write \@auxout {\string \citation {#2}}\fi
\@tempcntb\m@ne \let\@h@ld\relax \def\@citea{}%
\@cite{%
  \@for \@citeb:=#2\do {%
    \@ifundefined {b@\@citeb}%
      {\@h@ld\@citea\@tempcntb\m@ne{\bf ?}%
      \@warning {Citation `\@citeb ' on page \thepage \space undefined}}%
      {\@tempcnta\@tempcntb \advance\@tempcnta\@ne%
      \@tempcntb\number\csname b@\@citeb \endcsname \relax%
      \ifnum\@tempcnta=\@tempcntb %Number follows previous--hold on to it
        \ifx\@h@ld\relax%
          \edef \@h@ld{\@citea\csname b@\@citeb\endcsname}%
        \else%
          \edef\@h@ld{\ifmmode{-}\else--\fi\csname b@\@citeb\endcsname}%
        \fi%
      \else%   %  non-successor--dump what's held and do this one
        \@h@ld\@citea\csname b@\@citeb \endcsname%
        \let\@h@ld\relax%
      \fi}%
    \def\@citea{,\penalty\@highpenalty\,}%
  }\@h@ld
}{#1}}

\def\@citeb#1#2{{[#1]\if@tempswa , #2\fi}}
\def\@citeu#1#2{{$^{#1}$\if@tempswa , #2\fi }}
\def\@citep#1#2{{#1\if@tempswa , #2\fi}}

\def\bcites{         % cite with []'s
        \catcode`\@=11
        \let\@cite=\@citeb
        \catcode`\@=12
}

\def\upcites{         % cite with exponents
        \catcode`\@=11
        \let\@cite=\@citeu
        \catcode`\@=12
}

\def\plaincites{      % cite without brackets
        \catcode`\@=11
        \let\@cite=\@citep
        \catcode`\@=12
}

\newcount\hour
\newcount\minute
\newtoks\amorpm
\hour=\time\divide\hour by 60
\minute=\time{\multiply\hour by 60 \global\advance\minute by-\hour}
\edef\standardtime{{\ifnum\hour<12 \global\amorpm={am}%
        \else\global\amorpm={pm}\advance\hour by-12 \fi
        \ifnum\hour=0 \hour=12 \fi
        \number\hour:\ifnum\minute<10 0\fi\number\minute\the\amorpm}}
\edef\militarytime{\number\hour:\ifnum\minute<10 0\fi\number\minute}

\def\draftlabel#1{{\@bsphack\if@filesw {\let\thepage\relax
   \xdef\@gtempa{\write\@auxout{\string
      \newlabel{#1}{{\@currentlabel}{\thepage}}}}}\@gtempa
   \if@nobreak \ifvmode\nobreak\fi\fi\fi\@esphack}
        \gdef\@eqnlabel{#1}}
\def\@eqnlabel{}
\def\@vacuum{}
\def\marginnote#1{}
\def\draftmarginnote#1{\marginpar{\raggedright\scriptsize\tt#1}}
\def\draft{
        \pagestyle{plain}
        \overfullrule=2pt
        \oddsidemargin -.5truein
        \def\@oddhead{\sl \phantom{\today\quad\militarytime} \hfil
        \smash{\Large\sl DRAFT} \hfil \today\quad\militarytime}
        \let\@evenhead\@oddhead
        \let\label=\draftlabel
        \let\marginnote=\draftmarginnote
        \def\ps@empty{\let\@mkboth\@gobbletwo
        \def\@oddfoot{\hfil \smash{\Large\sl DRAFT} \hfil}
        \let\@evenfoot\@oddhead}
        \def\@eqnnum{(\theequation)\rlap{\kern\marginparsep\tt\@eqnlabel}%
        \global\let\@eqnlabel\@vacuum}  }

\def\blackfonts{
        \font\blackboard=msbm10 scaled\magstep1
        \font\blackboards=msbm8
        \font\blackboardss=msbm6
}

\def\prep{         % twocolumn.sty  Changed by Marek and Neil
        \catcode`\@=11
        \input art10.sty
        \catcode`\@=12
        
        \let\small\null
        \def\blackfonts{
                \font\blackboard=msbm10
                \font\blackboards=msbm7
                \font\blackboardss=msbm5
        }
        \let\sl\it
        \twocolumn
        \sloppy
        \voffset=-2.54truecm
        \hoffset=-2.54truecm
        \flushbottom
        \parindent 1em
        \leftmargini 2em
        \leftmarginv .5em
        \leftmarginvi .5em
        \marginparwidth 48pt
        \marginparsep 10pt
        \setlength{\columnsep}{2truecm}
        \setlength{\textwidth}{25.4truecm}
        \setlength{\textheight}{17truecm}
        \baselineskip=16pt
        \oddsidemargin .18truein
        \evensidemargin .17truein
}

\def\eqalign#1{\null\,\vcenter{\openup\jot\m@th
  \ialign{\strut\hfil$\displaystyle{##}$&$\displaystyle{{}##}$\hfil
      \crcr#1\crcr}}\,}
\def\eqalignno#1{\displ@y \tabskip\centering
  \halign to\displaywidth{\hfil$\@lign\displaystyle{##}$\tabskip\z@skip
    &$\@lign\displaystyle{{}##}$\hfil\tabskip\centering
    &\llap{$\@lign##$}\tabskip\z@skip\crcr
    #1\crcr}}

\def\section{\@startsection {section}{1}{\z@}{3.ex plus 1ex minus
 .2ex}{2.ex plus .2ex}{\large\bf}}
\def\subsection{\@startsection{subsection}{2}{\z@}{2.75ex plus 1ex minus
 .2ex}{1.5ex plus .2ex}{\bf}}        

\def\appendix{{\newpage\section*{Appendix}}\let\appendix\section%
        {\setcounter{section}{0}
        \gdef\thesection{\Alph{section}}}\section}

\def\abstract{\if@twocolumn
\section*{Abstract}
\else %\small
\begin{center}
{\bf Abstract\vspace{-.5em}\vspace{0pt}}
\end{center}
\quotation
\fi}

\catcode`\@=12
\def\d{\partial}

\def\sqr#1#2{{\vcenter{\vbox{\hrule height.#2pt\hbox{\vrule width.#2pt 
height#1pt \kern#1pt \vrule width.#2pt}\hrule height.#2pt}}}}

\def\eps{\varepsilon}

\def\=d{\,{\buildrel\rm def\over =}\,}

\def\N{\hbox{\bbf N}}

\def\i3p{\p32\int d^3p}

\def\As{A\hbox to 1pt{\hss /}}
\def\np4{\int d^4p_1\cdots d^4p_{n-1}\, }

\def\tr{{\rm tr}\, }

\def\nx4{\int d^4x_1\ldots d^4x_n\, }

\def\kon#1#2{\vbox{\halign{##&&##\cr
\lower4pt\hbox{$\scriptscriptstyle\vert$}\hrulefill &
\hrulefill\lower4pt\hbox{$\scriptscriptstyle\vert$}\cr $#1$&
$#2$\cr}}}

\def\konv#1#2#3{\hbox{\vrule height12pt depth-1pt}
\vbox{\hrule height12pt width#1cm depth-11.6pt}
\hbox{\vrule height6.5pt depth-0.5pt}
\vbox{\hrule height11pt width#2cm depth-10.6pt\kern5pt
      \hrule height6.5pt width#2cm depth-6.1pt}
\hbox{\vrule height12pt depth-1pt}
\vbox{\hrule height6.5pt width#3cm depth-6.1pt}
\hbox{\vrule height6.5pt depth-0.5pt}}
\def\konu#1#2#3{\hbox{\vrule height12pt depth-1pt}
\vbox{\hrule height1pt width#1cm depth-0.6pt}
\hbox{\vrule height12pt depth-6.5pt}
\vbox{\hrule height6pt width#2cm depth-5.6pt\kern5pt
      \hrule height1pt width#2cm depth-0.6pt}
\hbox{\vrule height12pt depth-6.5pt}
\vbox{\hrule height1pt width#3cm depth-0.6pt}
\hbox{\vrule height12pt depth-1pt}}

\def\konw#1#2#3{\hbox{\vrule height12pt depth-1pt}
\vbox{\hrule height12pt width#1cm depth-11.6pt}
\hbox{\vrule height6.5pt depth-0.5pt}
\vbox{\hrule height12pt width#2cm depth-11.6pt \kern5pt
      \hrule height6.5pt width#2cm depth-6.1pt}
\hbox{\vrule height6.5pt depth-0.5pt}
\vbox{\hrule height12pt width#3cm depth-11.6pt}
\hbox{\vrule height12pt depth-1pt}}

\def\i{{\rm int}}

\def\e{{\rm ext}}

\def\a{{\rm av}}

\def\m3{{\mu_1\mu_2\mu_3}}

\def\p{{(+)}}

\def\be{\begin{equation}}       \def\eq{\begin{equation}}
\def\ee{\end{equation}}         \def\eqe{\end{equation}}

\def\bea{\begin{eqnarray}}      \def\eqa{\begin{eqnarray}}
\def\ena{\end{eqnarray}}        \def\eea{\end{eqnarray}}
                                \def\eqae{\end{eqnarray}}

\def\ba{\begin{array}}
\def\ea{\end{array}}
\def\unit{1 \hskip-.3em \raise2pt\hbox{$ \scriptstyle |$ } }
\def\a{\alpha}
\def\b{\beta}
 
\def\d{\delta}
\def\e{\epsilon}           % Also, \varepsilon
\def\g{\gamma}
   
\def\i{\iota}

\def\l{\lambda}
\def\m{\mu}
\def\n{\nu}
  
\def\p{\pi}                % Also, \varpi
\def\t{\tau}

\def\D{\Delta}

\def\G{\Gamma}
\def\J{\Psi}

\def\bop#1{\setbox0=\hbox{$#1M$}\mkern1.5mu
        \vbox{\hrule height0pt depth.04\ht0
        \hbox{\vrule width.04\ht0 height.9\ht0 \kern.9\ht0
        \vrule width.04\ht0}\hrule height.04\ht0}\mkern1.5mu}
\def\Box{{\mathpalette\bop{}}}                        % box
\def\pa{\partial}                              % curly d
\def\>{\rangle} %right angle

\def\<{\langle} %left angle
\def\Dsl{D \hskip-.6em \raise1pt\hbox{$ / $ } }

\def\sl#1{\rlap{\hbox{$\mskip 1 mu /$}}#1}% good slash for l.c.
\def\leftrightarrowfill{$\mathsurround=0pt \mathord\leftarrow \mkern-6mu
       \cleaders\hbox{$\mkern-2mu \mathord- \mkern-2mu$}\hfill
       \mkern-6mu \mathord\rightarrow$}
\def\dvec#1{\vbox{\ialign{##\crcr
       \leftrightarrowfill\crcr\noalign{\kern-1pt\nointerlineskip}
       $\hfil\displaystyle{#1}\hfil$\crcr}}}          % <--> accent
\def\hook#1{{\vrule height#1pt width0.4pt depth0pt}}
\def\leftrighthookfill#1{$\mathsurround=0pt \mathord\hook#1
       \hrulefill\mathord\hook#1$}
\def\underhook#1{\vtop{\ialign{##\crcr                 % |_| under
       $\hfil\displaystyle{#1}\hfil$\crcr
       \noalign{\kern-1pt\nointerlineskip\vskip2pt}
       \leftrighthookfill5\crcr}}}
\def\smallunderhook#1{\vtop{\ialign{##\crcr      % " for su'scripts
       $\hfil\scriptstyle{#1}\hfil$\crcr
       \noalign{\kern-1pt\nointerlineskip\vskip2pt}
       \leftrighthookfill3\crcr}}}
\def\sfrac#1#2{{\vphantom1\smash{\lower.5ex\hbox{\small$#1$}}\over
       \vphantom1\smash{\raise.4ex\hbox{\small$#2$}}}} % alt. fraction
\def\bfrac#1#2{{\vphantom1\smash{\lower.5ex\hbox{$#1$}}\over
       \vphantom1\smash{\raise.3ex\hbox{$#2$}}}}      % "
\def\afrac#1#2{{\vphantom1\smash{\lower.5ex\hbox{$#1$}}\over#2}}  %"
\def\on#1#2{{\buildrel{\mkern2.5mu#1\mkern-2.5mu}\over{#2}}}%acc.over
\def\ddt#1{\on{\hbox{\LARGE .\kern-2pt.}}#1}             % double dot
\def\tdt#1{\on{\hbox{\LARGE .\kern-2pt.\kern-2pt.}}#1}   % triple dot

\def\boxes#1{
       \newcount\num
       \num=1
       \newdimen\downsy
       \downsy=-1.5ex
       \mskip-2.8mu
       \bo
       \loop
       \ifnum\num<#1
       \llap{\raise\num\downsy\hbox{$\bo$}}
       \advance\num by1
       \repeat}
\def\boxup#1#2{\newcount\numup
       \numup=#1
       \advance\numup by-1
       \newdimen\upsy
       \upsy=.75ex
       \mskip2.8mu
       \raise\numup\upsy\hbox{$#2$}}

\newskip\humongous \humongous=0pt plus 1000pt minus 1000pt
\def\caja{\mathsurround=0pt}
\def\eqalign#1{\,\vcenter{\openup2\jot \caja
       \ialign{\strut \hfil$\displaystyle{##}$&$
       \displaystyle{{}##}$\hfil\crcr#1\crcr}}\,}
\newif\ifdtup

\def\to{\rightarrow}

\def\1ov4{{1\over 4}}

\def\tr{{\rm tr}}

\def\pa{\partial}

\def\ddt{\dot{\t}}

\def\tr{\tilde{\r}}

\def\pa{\partial}

\renewcommand{\theequation}{\thesection.\arabic{equation}}

\def\appendix#1{
\addtocounter{section}{1}
\setcounter{equation}{0}
\renewcommand{\thesection}{\Alph{section}}
\section*{Appendix \thesection\protect\indent #1}
\addcontentsline{toc}{section}{Appendix \thesection\ \ \ #1}
}
\newcommand{\non}{\nonumber}

\newcommand{\la}{\langle}
\newcommand{\ra}{\rangle}

\def\pa{\partial}
\renewcommand{\a}{\alpha}
\renewcommand{\b}{\beta}

\renewcommand{\d}{\delta}
\newcommand{\5}{\tilde{5}}
\newcommand{\rmd}{{\rm d}}

\newcommand{\beq}{\begin{equation}}
\newcommand{\eeq}{\end{equation}}
\newcommand{\tD}{\tilde{\Delta}}
\def\ba{\begin{eqnarray}}
\def\ea{\end{eqnarray}}

\def\tr{{\rm tr}}
\def\N{{\cal N}}
\def\J{{\cal J}}
\def\Li{{\rm Li}_2}
\begin{document}

%%%%%%%%%

% Front page here

%\vspace*{1cm}
\null\vskip-24pt
\hfill LMU-TPW 00/12
\vskip-10pt
\hfill UAHEP 00/5
\vskip-10pt
\hfill KL-TH 00/04
\vskip-10pt
\hfill {\tt hep-th/0005182}
\vskip0.2truecm
\begin{center}
\vskip 0.2truecm
{\Large\bf
%\titleline
Operator Product Expansion of the Lowest Weight CPOs in ${\cal N}$=4 SYM$_4$
at Strong Coupling
}\\ 
\vskip 0.5truecm
%\vfill
{\bf
Gleb Arutyunov$^{*,**}$ 
\footnote{email:{\tt arut@theorie.physik.uni-muenchen.de }}, 
Sergey Frolov$^{\ddagger ,**}$ 
\footnote{email:{\tt frolov@bama.ua.edu}
%\newline

$^{**}$On leave of absence from 
Steklov Mathematical Institute, Gubkin str.8, 
117966, Moscow, Russia
}and 
Anastasios C. Petkou$^{\dagger}$
   \footnote{email:{\tt petkou@physik.uni-kl.de}
}  
}\\
\vskip 0.4truecm
%\addresses
$^{*}$
{\it Sektion Physik, Universit\"at M\"unchen,\\
Theresienstr. 37, D-80333 M\"unchen, Germany}\\
\vskip .2truecm
$^{\ddagger}$ {\it Department of Physics and Astronomy,
University of Alabama,\\ Box 870324,
Tuscaloosa, Alabama 35487-0324, USA}\\
\vskip .2truecm
$^{\dagger}${\it Department of Physics, Theoretical Physics,
University of Kaiserslautern,\\ Postfach 3049, 
67653 Kaiserslautern, Germany}\\
\end{center}
\vskip 0.2truecm
\noindent\centerline{\bf Abstract} 
We present a detailed analysis of the 4-point functions of the 
lowest weight chiral primary operators 
$O^{I} \sim \tr \phi^{(i}\phi^{j)}$ in
$\N =4$ SYM$_4$ at strong coupling and show 
that their structure is
compatible with  
the predictions of AdS/CFT correspondence. In particular,
all power-singular terms in the 4-point functions exactly
coincide with the contributions coming from the conformal blocks of
the CPOs, the R-symmetry current and the stress tensor. 
Operators dual to string modes decouple at strong
coupling. We compute the anomalous dimensions and the leading $1/N^2$
corrections to  the normalization constants 
of the 2- and 3-point functions
of scalar and vector double-trace operators with approximate
dimensions 4 and 5 respectively. We also find that the conformal
dimensions of certain towers of double-trace operators in the {\bf 105}, 
{\bf 84} and {\bf 175} irreps are non-renormalized. We show that,
despite the absence 
of a non-renormalization theorem for the double-trace operator in
the {\bf 20} irrep, its anomalous dimension vanishes. 
As by-products of our investigation, we derive explicit expressions
for the conformal block of the stress tensor, and for the 
conformal partial wave amplitudes of a conserved current and of
a stress tensor in $d$ dimensions. 

\newpage

%%%%%%%%%%%%%%%%%%%%%%%%%%%%%%%%%%%%%%%%%%%%%%%%%%%%%%%%%%%%%%%%%
\section{Introduction}
%%%%%%%%%%%%%%%%%%%%%%%%%%%%%%%%%%%%%%%%%%%%%%%%%%%%%%%%%%%%%%%%%

The AdS/CFT correspondence \cite{M,GKP,W} is arguably the best
currently available way of 
getting nontrivial dynamical information for the strong 
coupling behavior of certain conformal field theories.
In particular, the $\N =4$ supersymmetric $SU(N)$ Yang-Mills 
theory in four dimensions (SYM$_4$) at large $N$ and at strong `t Hooft
coupling $\lambda =g_{YM}^2 N$ is dual to type IIB supergravity on
the $AdS_5\times S^5$ background. The supergravity fields are dual
to certain quasi-primary operators in SYM$_4$. 
According to \cite{GKP,W}, the generating functional for the  connected
Green functions of these operators coincides with the on-shell value
of type IIB supergravity action which has to be further modified by 
the addition of 
definite boundary terms \cite{AF1}. Thus, computing
$n$-point correlation functions in the supergravity approximation 
is generally divided into two independent problems - 
finding first the supergravity action up to the $n$-th order and 
then evaluating its on-shell value.
Although a covariant action for
type IIB supergravity is unknown, one can use the covariant equations 
of motion \cite{S,SW,HW} and the quadratic action \cite{AF3} to find
cubic actions \cite{LMRS,AF5,Lee} for its physical fields  and to
compute the corresponding 3-point functions 
using the technique developed in \cite{FMMR}. 

Computing 4-point functions \cite{LT1}-\cite{HPR2} in the 
supergravity approximation in general
requires the derivation of the supergravity action up to 
fourth order. The 
part of the action relevant to the massless modes, corresponding to the dilaton
and axion fields, was already known as pointed out in \cite{LT1}
where the calculation of the  
corresponding 4-point functions was initiated. The complete expression 
for the 4-point functions was obtained in \cite{HFMMR} and was 
further analyzed in \cite{HMMR}. Unfortunately, these modes are dual
to the
rather complicated operators $\tr F^2 +...$ and $\tr F\tilde{F}$ 
and the analysis performed in \cite{HMMR} was
unavoidably incomplete. 

It is known that all operators dual to the type IIB supergravity
fields belong to short representations of the conformal superalgebra
$SU(2,2|4)$ and are supersymmetric descendents of Chiral Primary
Operators (CPOs) of the form 
$O_k^I=\tr (\phi^{(i_1}\cdots\phi^{i_k)})$. CPOs are dual to scalar
fields $s^I$ that are mixtures of the five form field strength on $S^5$
and the trace of the graviton on $S^5$. The relevant part of the 
quartic action of type IIB supergravity for the scalars $s^I$ was found in
\cite{AF6} and was then used in \cite{AF7} to compute 
the 4-point functions of the simplest CPOs 
$O^I=\tr (\phi^{(i}\phi^{j)})$. In the present  paper we use 
these 4-point functions to analyze in detail the 
Operator Product Expansion (OPE) of the 
lowest weight CPOs at strong coupling.  

It is widely believed that the structure of a Conformal Field Theory (CFT) is
encoded in the 
OPE since knowledge of the latter allows, in principle, the
calculation of all $n$-point functions. 
Thus, in the context of AdS/CFT correspondence one would eventually
like to prove that the 4-point functions, (and in general $n$-point
functions) of CPOs in the boundary CFT  computed in the supergravity
approximation admit an OPE interpretation. This  is a rather
complicated problem
because an infinite number of quasi-primary operators
may in principle appear in the OPE of two CPOs. Therefore, the best
one can presently do is to show that 
the leading terms in a double OPE expansion of the 4-point functions
exactly match the contributions of the conformal 
blocks of the first few quasi-primary operators
with the lowest conformal dimensions. This is the main line of investigation
which we follow in the present work in our analysis of the
4-point function of the lowest weight CPOs in ${\cal N}=4$
SYM$_4$. 

Our study 
shows that there are four  
singular terms in the OPE of two lowest weight CPOs 
corresponding to the identity operator, the lowest weight CPO itself,
the $R$-symmetry  
vector current and the stress tensor. These  three nontrivial operators
are dual to the scalars $s^I$, the vector fields $A_\mu$ and 
the graviton $h_{\mu\nu}$ that appear in the exchange Feynman diagrams
of type IIB supergravity. The most singular terms in the 4-point
functions computed in the supergravity approximation exactly coincide
with the contributions coming from the conformal blocks of the above three
operators. 

We compare the strong coupling OPE with the free field theory OPE, and 
explicitly  observe, at weak coupling,  the splitting of the $R$-symmetry
current and of the stress 
tensor into 2 and 3 terms respectively which belong to different
supermultiplets. Only one term in each splitting is 
dual to a supergravity field and survives  at strong
coupling while the others acquire large dimensions and decouple. 
A similar type of splitting also occurs in the case of the double-trace 
operators transforming in the ${\bf 84}$ and ${\bf 175}$ irreps. 
 
We also analyze the leading nonsingular terms in the OPE which are  
due to double-trace operators of the schematic form $:\pa^m O^I \pa^n
O^J:$ with free field conformal dimensions $4+m+n$.
A  generic property of any correlation function
computed in the supergravity approximation is the appearance of 
logarithmic terms. In an unitary CFT logarithmic terms have a natural
interpretation in terms of anomalous dimensions of 
operators \cite{Sym} and such an interpretation was used 
in the past in studies of the $O(N)$ 
vector model \cite{LR,P}. Since the operators dual to the supergravity
fields have protected conformal dimensions, the logarithmic terms in 
the  correlation functions of supergravity can only be attributed to 
anomalous dimensions of double-trace operators. 

We show that among the scalar double-trace operators 
with free field conformal dimension 4, 
the only one  
acquiring an anomalous dimension is the operator
$:O^IO^I:$, which transforms in the 
trivial representation of the
$R$-symmetry group 
$SO(6) \sim SU(4)$. The anomalous dimension of this operator is found to be
$-16/N^2$ and coincides with the anomalous dimension of the 
operator $B$ which was calculated in \cite{HMMR}. This is consistent
with the fact that $B$ is a supersymmetric descendent of $:O^IO^I:$. 
It is worth noting that among the non-renormalized operators we find a
double-trace scalar operator in the ${\bf 20}$ irrep of
$SO(6)$ whose non-renormalization property does not follow from the
shortening condition discussed in \cite{AFer,AFer2}. 

Finally, we compute the anomalous dimensions of 
the double-trace vector operators with free field conformal dimension 5
transforming in the ${\bf 15}$ and ${\bf 175}$ irreps respectively. 
We show that there are several towers of traceless symmetric 
tensor operators in the {\bf 105}, 
{\bf 84} and {\bf 175} irreps, whose anomalous conformal dimensions 
vanish. Some of these tensor operators are not subject to 
any known non-renormalization theorem.

The 4-point functions of CPOs also allow us to find the leading 
$1/N^2$ corrections to the normalization constants of the 3-point
functions involving two CPOs and one double-trace operator with
low conformal dimension. 
In the case when a double trace operator has protected dimension 
we interpret these corrections  
as manifestation of the splitting of the free field theory 
operator in two orthogonal parts carrying different representations
of supersymmetry. The first one has protected both the dimension 
and the normalization constant, the other one acquires infinite 
anomalous dimension and disappears at strong coupling. 
To make this interpretation precise one should further show  that 
the linear splitting arising due to the difference between 
normalization constants in free theory and at strong coupling is consistent 
with the fact that the split fields transform in different
representations of supersymmetry. It would be quite interesting to  
investigate such a property in more detail.  
 
The plan of the paper is as follows.
In section 2 we recall how logarithmic terms are related to 
anomalous conformal dimensions in an unitary CFT and in the framework
of the AdS/CFT correspondence. In section 3 we discuss the structure of
the OPE of the lowest weight CPOs in free field theory and at
strong coupling. In section 4 we compute anomalous dimensions and 
first corrections to the 2- and 3-point normalization 
constants of double-trace operators of approximate dimensions 4 and 5.
A discussion of the results obtained and our conclusions are
presented in section 5. Several technical issues are considered in
five Appendices. In the Appendix A we discuss a decomposition of
a bi-local operator which is a normal-ordered product of two 
quasi-primary scalar operators into a sum of conformal blocks
of local tensor primary operators. In the Appendix B explicit 
formulae for conformal partial amplitudes of scalar, conserved vector
current and stress tensor are derived. A convenient series representation
used throughout the paper is obtained in the Appendix C.
In the Appendix D we discuss the projectors which single out the 
contributions of irreps occurring in the decomposition 
${\bf 20}\times{\bf 20}$ of $SO(6)$ from the 4-point function of CPOs.
In the Appendix E an explicit formula for the conformal block of the stress
tensor is derived.    
%%%%%%%%%%%%%%%%%%%%%%%%%%%%%%%%%%%%%%%%%%%%%%%%%%%%%%%%%%%%%%%%%%%%%
\section{Anomalous dimensions and logarithmic terms in CFT}
\setcounter{equation}{0}
%%%%%%%%%%%%%%%%%%%%%%%%%%%%%%%%%%%%%%%%%%%%%%%%%%%%%%%%%%%%%%%%%%%%%
An arbitrary unitary CFT is completely 
characterized by a set of quasi-primary  
operators $O_i$ of conformal dimensions $\D_i$ and by their OPE
\ba
O_i(x)O_j(y) = \sum_k\, \frac{1}{|x-y|^{\D_i+\D_j-\D_k}}\,
C_{ij}^k(x-y,\pa_y )O_k(y)\,.
\label{1}
\ea
Here the sum runs over the set of all the quasi-primary operators  and
$i,j,k$ are 
multi-indices which in general include the indices of the $R$-symmetry 
and of the Lorentz groups. The operator algebra structure constants 
$C_{ij}^k(x-y,\pa_y )$ can be decomposed in a power series in $x-y$
and $\pa_y$.  
Without loss of generality one can assume
that the operators $O_i$ are orthogonal
\ba
\la O_i(x)O_j(0)\ra =C_i\frac{\d_{ij}}{x^{2\D_i}},
\non
\ea
where $C_i$ is a normalization constant of the 2-point function.
Then the operator algebra structure constants are fixed by the 
conformal dimensions $\D_i,\D_j,\D_k$, and by the ratio  
$C_{ijk}/C_k$, where the structure constants
$C_{ijk}$ appear in the 3-point functions
\ba
\la O_i(x)O_j(y)O_k(z)\ra =
\frac{C_{ijk}}{|x-y|^{\D_i+\D_j-\D_k}|x-z|^{\D_i+\D_k-\D_j}
|y-z|^{\D_j+\D_k-\D_i}}.
\label{3}
\ea 
The conformal dimensions and the structure constants depend on
the coupling constants of the CFT. 
In principle, the OPE (\ref{1}) allows one to 
compute any correlation function in the CFT. In particular, 
4-point functions are given by the following (schematic) double OPE expansion
\ba
\la O_i(x)O_j(y)O_k(z)O_l(w)\ra &=& \sum_{m}\,
\frac{1}{|x-y|^{\D_i+\D_j-\D_m}|z-w|^{\D_k+\D_l-\D_m}}\nonumber \\
&~&\quad\times\, C_{ij}^m(x-y,\pa_y )
\,C_{kl}^m(z-w,\pa_w)\frac{C_m}{|y-w|^{2\D_m}}.
\label{4}
\ea 
Thus we see that the short distance expansion of exact CFT correlation
functions does not contain logarithmic terms. 
Suppose, however, that one can only calculate correlation functions up to
some order in the coupling constant or another small parameter of the CFT.  
Then it is clear from (\ref{4}) that logarithmic terms would appear
due to the nontrivial dependence of conformal dimensions on the 
coupling or on the small parameter. These terms  can be easily found 
representing the conformal dimensions as  
$\D = \D^{(0)} +\D^{(1)}$, where $\D^{(0)}$ is the ``canonical'' part
and $\D^{(1)}$ is the ``anomalous'' coupling constant dependent
  part. Such a representation leads then to an expansion for the
  two-point functions of the form 
$|x|^{\D^{(1)}}=1+\D^{(1)}\log |x| +...$, connecting the
logarithmic  
terms to the anomalous dimensions, that
may be used to compute the latter. It is worthwhile to note that at the $n$-th
order of perturbation theory one encounters terms of
the form $(\log |x|)^n$.  

The $\N =4$ SYM$_4$ theory provides an example of such a logarithmic
behavior of  
correlation functions, both in the weak coupling standard perturbation
expansion \cite{EHSSW}-\cite{BKRS2} and also 
in the supergravity approximation \cite{HFMMR,AF7}. 
Due to superconformal invariance all quasi-primary operators of SYM$_4$ 
belong either to short or long representations of the conformal
superalgebra $SU(2,2|4)$ and in the framework of the AdS/CFT correspondence
fall into three classes:

\noindent {\it i}) Chiral operators dual to the 
type IIB supergravity fields which belong
to short representations and have protected conformal dimensions.
The simplest operators in this  class are the lowest weight CPOs 
$O^I=\tr (\phi^{(i}\phi^{j)})$.

\noindent {\it ii}) Operators dual to multi-particle 
supergravity states which are 
obtained as ``normal-ordered'' products of the chiral operators,
e.g. the double-trace operators $:O^IO^J:$. They may belong  
either to short or long representations and have conformal dimensions
restricted from above.

\noindent {\it iii}) Operators dual to string states 
(single- or multi-particle)
which belong to long representations and whose conformal dimensions
grow as $\lambda^{1/4}$ in the strong coupling limit. The simplest
example of such an operator is the Konishi operator $\tr (\phi^i\phi^i)$.
 
In the supergravity approximation to the AdS/CFT correspondence the
operators dual to string states decouple from the spectrum and
one can calculate the connected $n$-point functions of chiral operators
dual to the supergravity fields to leading order which is $1/N^{n-2}$.  
Since the expansion parameter is $1/N^2$, an $n$-point function 
contains logarithmic terms of the form $(\log |x|)^{[(n-2)/2]}$. 
In particular, a 4-point function can have only $\log |x|$-dependent terms,
and cannot have, say, terms of the form $(\log |x|)^2$. 
Moreover, since chiral  operators have protected conformal dimensions
only the operators dual to multi-particle supergravity states 
contribute to $\log$-dependent terms. 

The AdS/CFT correspondence 
predicts a simple form of the OPE of chiral operators in the strong coupling
limit. Let $O_1$ and $O_2$ be operators dual to the supergravity fields
$\varphi_1$ and $\varphi_2$ respectively and let the supergravity action
contain the non-vanishing cubic couplings
$\frac{1}{N}\lambda_{12k}\varphi_1\varphi_2\varphi_k$ with 
some fields  $\varphi_k$. 
Then, the OPE of $O_1$ and $O_2$ takes the form (suppressing 
the indices of the operators and structure constants) 
\ba
O_1(x)O_2(y) = \frac{1}{N}\sum_k\, \frac{1}{|x-y|^{\D_1+\D_2-\D_k}}\,
C_{12}^k(x-y,\pa_y )O_k(y) + [:O_1(x)O_2(y):],
\label{ope12}
\ea
where $O_k$ is an operator dual to $\varphi_k$.
Here we denote by  $[:O_1(x)O_2(y):]$ an infinite sum of tensor 
quasi-primary operators and their descendents, which are dual to multi-particle
supergravity states. In general these operators acquire anomalous
dimensions and are responsible for the appearance of logarithms in
correlation functions. An important property of the operators dual to
multi-particle supergravity states is
that their structure constants are of order 1, while the structure
constants of the operators dual to supergravity fields are of order $1/N$.
Due to such a property,
the sum of these operators coincides in the limit $N\to\infty$ with
the corresponding free field theory  
normal-ordered operator $:O_1^{fr}(x)O_2^{fr}(y):$. This can be seen
as follows. 
A 4-point function of chiral operators is given by a sum of a disconnected
contribution which is of order 1 and a connected Green function
which is of order $1/N^2$. Since the structure constants of the 
operators dual to supergravity fields are of order $1/N$, they do not 
contribute to the disconnected part of the 4-point function. Thus 
only the ``normal-ordered'' operators contribute. The disconnected
part is given by a sum of products of 2-point functions of chiral
operators, hence it does not depend on the coupling constant and $N$
(we assume that all the chiral operators are orthonormal) and 
coincides with the free field disconnected part. Therefore, in the limit
$N\to\infty$ the sum \mbox{$[:O_1(x)O_2(y):]$} has to coincide with the 
free field normal-ordered product  $:O_1^{fr}(x)O_2^{fr}(y):$,\footnote{
One can easily see that the normal-ordered product $:O_1^{fr}(x)O_2^{fr}(y):$
is the only term of order 1 in the free field OPE of chiral operators.} 
that is decomposed into a sum of local tensor quasi-primary operators. 
However, at finite $N$ an infinite number of the tensor operators 
acquire anomalous dimensions and their structure constants get 
$1/N^2$ corrections to their free field values. For this reason it
seems hardly possible to prove that a 4-point function computed in
the supergravity approximation admits an OPE interpretation.   
This would require the knowledge of the conformal partial wave amplitude of
an arbitrary tensor operator. Another reason that complicates  the analysis
of 4-point functions is that in general one should split
the free field theory double-trace operators into a sum of
operators with the same free field theory dimensions, each one
transforming irreducibly under the superconformal group. 
In the context of the present work we are able to successfully deal with
both the above problems.
%%%%%%%%%%%%%%%%%%%%%%%%%%%%%%%%%%%%%%%%%%%%%%%%%%%%%%%%%%%%%%%%%%%
\section{OPE of the lowest weight CPOs}
\setcounter{equation}{0}
%%%%%%%%%%%%%%%%%%%%%%%%%%%%%%%%%%%%%%%%%%%%%%%%%%%%%%%%%%%%%%%%%%%%%
In this section we study the OPE of the lowest weight CPOs in
free field theory and at strong coupling.  
Recall that the normalized lowest weight CPOs in ${\cal N}=4$
SYM$_4$ are operators of the form 
$$
O^I(x)=\frac{2^{3/2}\pi^2 }{\lambda}C_{ij}^I\tr(:\phi^i\phi^j:),
$$
where the symmetric traceless  
tensors $C_{ij}^I$, $i,j=1,2,..,6$ form a basis of the 
${\bf 20}$ of $SO(6)$ and satisfy the orthonormality condition
$$C_{ij}^IC_{ij}^J=\d^{IJ}.$$
Using for the Wick contractions the following propagator 
\ba
\langle \phi_a^i\phi^j_b \rangle = 
\frac{g_{YM}^2\d_{ab}\d^{ij}}{(2\pi)^2x_{12}^2},
\label{prop}
\ea
where $a,b$ are color indices and 
$x_{ij}=x_i-x_j$, one finds 
the following 
expressions for the free field theory 
2-, 3- \cite{LMRS} and 4-point
functions of $O^I$: 
\ba
&&\langle O^{I_1}(x_1)O^{I_2}(x_2)  
\rangle_{fr} =\frac{\d^{I_1I_2}}{x_{12}^2},
\nonumber \\
&&\langle O^{I_1}(x_1)O^{I_2}(x_2)O^{I_3}(x_3) 
\rangle_{fr}  =\frac{1}{N}\frac{2^{3/2}
C^{I_1I_2I_3}}{x_{12}^2x_{13}^2
  x_{23}^2},  
\nonumber \\
&&\langle O^{I_1}(x_1)O^{I_2}(x_2)O^{I_3}(x_3)O^{I_4}(x_4)\rangle _{fr} =  
\Biggl[\frac{\d^{I_1I_2}\d^{I_3I_4}}{x_{12}^4x_{34}^4}
+\frac{\d^{I_1I_3}\d^{I_2I_4}}{x_{13}^4x_{24}^4}
+\frac{\d^{I_1I_4}\d^{I_2I_3}}{x_{14}^4x_{23}^4}\Biggl]\nonumber \\
&&\hspace{3cm} 
+\frac{4}{N^2}\Biggl[
 \frac{C^{I_1I_2I_3I_4}}{x_{12}^2x_{23}^2 
   x_{34}^2 x_{41}^2} +\frac{C^{I_1I_3I_2I_4}}{x_{13}^2x_{32}^2
   x_{24}^2 x_{41}^2}+\frac{C^{I_1I_3I_4I_2}}{x_{13}^2x_{34}^2
   x_{42}^2 x_{21}^2}\Biggl],
\label{free4}
\ea
where the first term in the 4-point function represents the contribution 
of disconnected diagrams. We have also introduced the shorthand 
notations $C^{I_1I_2I_3}=C^{I_1}_{i_1i_2}C^{I_2}_{i_2i_3}
C^{I_3}_{i_3i_1}$ and
$C^{I_1I_2I_3I_4}=C^{I_1}_{i_1i_2}C^{I_2}_{i_2i_3}
C^{I_3}_{i_3i_4}C^{I_4}_{i_4i_1}$
for the trace products of matrices $C^{I}$.
%%%%%%%%%%%%%%%%%%%%%%%%%%%%%%%%%%%%%%%%%%%%%%%%%%%%%%%%%%%%%%%%%
\subsection{Free field theory OPE}
%%%%%%%%%%%%%%%%%%%%%%%%%%%%%%%%%%%%%%%%%%%%%%%%%%%%%%%%%%%%%%%
The simplest way to derive the OPE in free field theory is to 
apply Wick's theorem. Using the propagator (\ref{prop}) 
we find the following formula for the product of two CPOs
\ba
O^{I_1}(x_1)O^{I_2}(x_2)&=&\frac{\d^{I_1I_2}}{x_{12}^4}
+\frac{2^3\pi^2}{\lambda N x_{12}^2}C_{ik}^{I_1}C_{kj}^{I_2}:\tr
(\phi^i(x_1)\phi^j(x_2)): \nonumber \\
&+&:O^{I_1}(x_1)O^{I_2}(x_2):
\label{basic} 
\ea
On the r.h.s. of (\ref{basic}) we have bi-local operators of the form
$:O^\alpha (x_1)O^\beta (x_2):$, where $O^\alpha$ is either
$\phi^i$ or $O^{I_1}$ and  $O^\beta$ is either
$\phi^j$ or $O^{I_2}$.
To find the operator content of the r.h.s. of (\ref{basic}) one should
perform the Taylor expansion of the operator $O^\alpha$ 
and rearrange the resulting series as
a sum of conformal blocks of local quasi-primary operators.
It is clear that in free field theory any bilocal operator 
$:O^\alpha (x_1)O^\beta (x_2):$
may be represented as an infinite sum of conformal blocks of symmetric 
traceless rank $l$ tensor operators with dimensions 
$\D_\alpha +\D_\beta + l +2k$, 
\ba
:O^\alpha (x)O^\beta (0):=
\sum_{l,k=0}^\infty\, 
\frac{1}{(l+2k)!}x^{2k}\,x^\mu_1\cdots x^\mu_l
 [O^{(k)}_{\mu_1\cdots \mu_l}(0)]\, ,
\label{biloc}
\ea
where the square brackets $[~]$ are used 
to denote the whole conformal block of a quasi-primary operator. 
In an interacting theory the tensor quasi-primary operators
may acquire anomalous dimensions.
Explicit expressions of the tensor operators
through $O^\alpha ,O^\beta$ are unknown
 and the best we can do is 
to find the first few terms in the series. In particular, as shown in 
Appendix A, the terms up to two derivatives are given by
the following formula
\ba
:O^\alpha (x)O^\beta (0):&=& :O^\alpha (0)O^\beta (0):+
x^\mu:\pa_\mu O^\alpha (0)O^\beta (0):+
\frac12  x^\mu x^\nu :\pa_\mu\pa_\nu O^\alpha (0)O^\beta (0):\cr
&=&[O^{\alpha\beta }(0)] + x^\mu [O_\mu^{\alpha\beta}(0)] -
\frac12 x^\mu x^\nu [T_{\mu\nu}^{\alpha\beta}(0)] +
\frac12 x^2 [T^{\alpha\beta}(0)].
\label{taylor}
\ea
Here the quasi-primary operators are given by
\ba
O^{\alpha\beta} &=& :O^{\alpha}O^{\beta}:\, , \cr
O_\mu^{\alpha\beta} &=& \frac 12 :(\pa_\mu O^{\alpha}O^{\beta}-
O^{\alpha}\pa_\mu O^{\beta}): \, , \nonumber \\
T_{\mu\nu}^{\alpha\beta}&=&\frac 12 :(\pa_\mu O^{\alpha}\pa_\nu O^{\beta}+
\pa_\nu O^{\alpha}\pa_\mu O^{\beta}):
-\frac{\D}{2(2\D +1)}\pa_\mu\pa_\nu (:O^{\alpha}O^{\beta}:)\nonumber \\
&+&\frac{\d_{\mu\nu}}{8}\left( 
-\frac{\D +1}{2\D +1}\pa^2 (:O^{\alpha}O^{\beta}:)+
:\pa^2O^{\alpha}O^{\beta}:+:O^{\alpha}\pa^2O^{\beta}:\right) \, ,
\nonumber \\
T^{\alpha\beta}&=&\frac{1}{8}\left( 
-\frac{\D -1}{2\D -1}\pa^2 (:O^{\alpha}O^{\beta}:)+
:\pa^2O^{\alpha}O^{\beta}:+:O^{\alpha}\pa^2O^{\beta}:\right) \, ,
\nonumber
\ea
where $\D$ is the conformal dimension of the operators 
$O^\alpha ,O^\beta$ which takes the values 1 and 2  in the cases under
consideration.

Obviously the conformal dimensions of the scalar operators
$O^{\alpha\beta}$ and $T^{\alpha\beta}$ are equal to $2\D$ and $2\D +2$
respectively, the dimension of the 
vector operator is $2\D +1$ and the dimension of the traceless 
symmetric tensor operator is $2\D +2$. Consider first the case when 
$\D =1$. The scalar operator $\tr (\phi^i\phi^j)$ is decomposed
into a sum of the traceless part in the {\bf 20} - which is a lowest
weight CPO $O^I$ - 
and the trace part. The trace part is the normalized Konishi scalar field 
${\cal K}=\frac{2\pi^2}{3^{1/2}\lambda}\tr (\phi^2)$.
If $\D =1$ the vector and tensor
operators are conserved and  the operator 
$T^{ij}$ vanishes because of the on-shell equation $\pa^2\phi^i =0$. 
In fact the conserved current transforms in the ${\bf 15}$ irrep of 
$SO(6)$ and is the $R$-symmetry current of the 
free field theory of 6 scalars $\phi^i$. Decomposing the tensor
operator $T_{\mu\nu}^{ij}$ into irreducible representations of
the $R$-symmetry group $SO(6)$, i.e. into the traceless and 
trace parts with respect to the indices $i,j$, one sees 
that the trace part $T_{\mu\nu}^{ii}$
coincides with the stress tensor of the free field theory. 
The Konishi scalar and the
traceless part of $T_{\mu\nu}^{ij}$ are dual to string modes 
and are expected to decouple in the strong coupling limit.

To complete the consideration of the free field theory OPE we 
have to decompose the remaining operators into irreducible
representations of $SO(6)\sim SU(4)$. 
One has the general decomposition of the $\bf{20}\times \bf{20}$
of $SU(4)$ as
\ba
\bf{20}\times \bf{20}&=&[0,0,0]+[0,2,0]+[0,4,0]
+[2,0,2]\nonumber \\
&+&[1,0,1]+[1,2,1].
\label{irreps}
\ea
The representations in the first and the second 
lines of (\ref{irreps}) are symmetric and antisymmetric in
the indices of the ${\bf 20}$'s $I_1, I_2$, respectively.
The dimensions of the representations are 
\ba
&&D([0,0,0])=1,\quad D([0,2,0])=20,\quad D([0,4,0])=105,\quad
D([2,0,2])=84,\nonumber \\
&&D([1,0,1])=15, \quad D([1,2,1])=175 .
\label{dimreps}
\ea
Introducing the orthonormal Clebsh-Gordon coefficients 
$C^{I_1I_2}_{{\cal J}_D}$
$$C^{I_1I_2}_{{\cal J}_{D}}C^{I_1I_2}_{{\cal J}'_{D}}=
\d_{{\cal J}_{D}{\cal J}'_{D}},$$
where ${\cal J}_D$
is the index of an irrep of dimension $D$, as well as the operators
\ba
O^{{\cal J}_D}=C^{I_1I_2}_{{\cal J}_D}:O^{I_1}O^{I_2}:,\quad
O_\mu^{{\cal J}_D}=C^{I_1I_2}_{{\cal J}_D}O_\mu^{I_1I_2},
\label{doubtr}
\ea
we can write
\ba
:O^{I_1}O^{I_2}:&=&\d^{I_1I_2}O_{1}+
C^{I_1I_2}_{{\cal J}_{20}}O^{{\cal J}_{20}}
+C^{I_1I_2}_{{\cal J}_{105}}O^{{\cal J}_{105}}+
C^{I_1I_2}_{{\cal J}_{84}}O^{{\cal J}_{84}}\nonumber \\
O_\mu^{I_1I_2}&=&\frac12 \left( :\pa_\mu O^{I_1}O^{I_2}:
-:O^{I_1}\pa_\mu O^{I_2}:\right) 
=C^{I_1I_2}_{{\cal J}_{15}}O_\mu^{{\cal J}_{15}}
+C^{I_1I_2}_{{\cal J}_{175}}O_\mu^{{\cal J}_{175}},
\nonumber
\ea
and a similar decomposition for $T_{\mu\nu}^{I_1I_2}$ and 
$T^{I_1I_2}$. Note that the operators have the following free field 
theory 2-point 
functions\footnote{The only exception is the operator 
$O_1=\frac{1}{20}:O^{I_1}O^{I_2}:$ in the 
singlet representation, whose normalization constant is 
$\frac{1}{10} +O(\frac{1}{N^2})$.}
\ba
&&\langle O^{\J_1}(x_1)O^{\J_2}(x_2)  \rangle 
=\Bigl(2 +O(\frac{1}{N^2})\Bigl)\frac{\d^{\J_1\J_2}}{x_{12}^8}\, ,
\nonumber  \\
&&\langle O_\mu^{\J_1}(x_1)O_\nu^{\J_2}(x_2) 
\rangle =\Bigl(4 +O(\frac{1}{N^2})\Bigl)
\frac{I_{\mu\nu}(x_{12})}{x_{12}^{10}}\d^{\J_1\J_2}\, , 
\nonumber
\ea
where $I_{\mu\nu}(x)=\d_{\mu\nu}-2\frac{x_\mu x_\nu}{x^2}$.
The precise values of the normalization constants will be determined
in the next section. Due to the definition of the double-trace
operators, the 3-point normalization constants which appear 
in the following 3-point functions
\ba
\la O^{I_1}(x_1)O^{I_2}(x_2)O^{\J_D}(x_3)\ra &=&C_{OOO_{\bf D}}
\frac{C^{I_1I_2}_{{\cal J}_{D}}}
{|x_{12}|^{4-\D_D}|x_{13}|^{\D_D}|x_{23}|^{\D_D}} \, ,\nonumber \\
\la O^{I_1}(x_1)O^{I_2}(x_2)O_\mu^{\J_D}(x_3)\ra &=&C_{OOO_{\bf D}}
\frac{C^{I_1I_2}_{{\cal J}_{D}}\left(x_{23}^2x_{31}^\mu-x_{31}^2x_{23}^\mu
\right)}
{|x_{12}|^{5-\D_D}|x_{13}|^{\D_D +1}|x_{23}|^{\D_D +1}}
\nonumber
\ea 
are equal to the 2-point normalization constants $C_{O_{\bf D}}$.

Combining all pieces together we obtain the first few terms
in the free field OPE of the CPOs as
\ba
O^{I_1}(x_1)O^{I_2}(x_2)&=&\frac{\d^{I_1I_2}}{x_{12}^4}
+\frac{2^{3/2}}{N}C^{I_1I_2I}\frac{1}{x_{12}^2}[\,O^I\,]
+\frac{2}{3^{1/2} N}\d^{I_1I_2}
\frac{1}{x_{12}^2}[\,{\cal K}\,] \nonumber \\
&&\hspace{-1cm}+\, \frac{2^{7/2}\pi^2}{\lambda N}\frac{x_{12}^{\mu}}{x_{12}^2}
C^{I_1I_2}_{\J_{15}}[\,J^{\J_{15}}_{\mu}\,] 
-\frac{2\pi^2\d^{I_1I_2}}{3\lambda N }
\frac{x_{12}^{\mu}x_{12}^{\nu}}{x_{12}^2}[\,T_{\mu\nu}^{fr}\,]
+\frac{4\pi^2}{\lambda N}
\frac{x_{12}^{\mu}x_{12}^{\nu}}{x_{12}^2}C^{I_1I_2I}
[\,T_{\mu\nu}^{I}\,]\nonumber \\ 
&&\hspace{-1cm}+\,\d^{I_1I_2}[\,O_{1}\,]+
C^{I_1I_2}_{{\cal J}_{20}}[\,O^{{\cal J}_{20}}\,]
+C^{I_1I_2}_{{\cal J}_{105}}[\,O^{{\cal J}_{105}}\,]+
C^{I_1I_2}_{{\cal J}_{84}}[\,O^{{\cal J}_{84}}\,]\nonumber \\
&&\hspace{-1cm}+\, C^{I_1I_2}_{{\cal J}_{15}}x^\mu_{12}[\,O_\mu^{{\cal
      J}_{15}}\,] 
+C^{I_1I_2}_{{\cal J}_{175}}x^\mu_{12}[\,O_\mu^{{\cal J}_{175}}\,]+\ldots .
\label{freeope}
\ea
Here $T_{\mu\nu}^{fr}$ is the stress tensor of the 
free field theory of six scalar
fields, while the normalized $R$-symmetry current $J^{\J_{15}}_{\mu}$ 
is defined as follows
$$   
J^{\J_{15}}_{\mu}= 
C^{\J_{15}}_{ij}\frac12 \tr\left( :\pa_\mu \phi^i\phi^j:
-:\phi^i\pa_\mu \phi^j:\right) ,
$$
where the antisymmetric tensors $C^{\J_{15}}_{ij}$ form a basis of
the ${\bf 15}$ of $SO(6)$ and satisfy the orthogonality condition
$C^{\J_{15}}_{ij}C^{\J_{15}'}_{ij}=\d^{\J_{15}\J'_{15}}$. 
The $R$-symmetry current has the following 2-point function
$$
\langle J_\mu^{\J_{15}}(x_1)J_\nu^{\J_{15}'}(x_2) 
\rangle =
\frac{\lambda^2}{8\pi^4}
\d^{\J_{15}\J_{15}'}\frac{I_{\mu\nu}(x_{12})}{x_{12}^{6}}.
$$
We would like to stress that in addition to the above
fields the OPE contains infinite towers of
both single-trace as well as double-trace operators.
%%%%%%%%%%%%%%%%%%%%%%%%%%%%%%%%%%%%%%%%%%%%%%%%%%%%%%%%%%%%%%%%%%%%%%%%%%%%%%
\subsection{Strong coupling OPE}
%%%%%%%%%%%%%%%%%%%%%%%%%%%%%%%%%%%%%%%%%%%%%%%%%%%%%%%%%%%%%%%%%%%%%%%%%%%%%%
As was explained in the previous section, the strong coupling OPE
of CPOs is easily determined from the cubic terms
in the scalars $s^I$  dual to the lowest weight CPOs  
in the type IIB supergravity action. There are three different 
cubic vertices in the action describing the cubic
couplings among the three scalars $s^I$, the interaction of the  
scalars with the graviton and the interaction with 
the $SO(6)$ vector fields. Thus, according to the discussion in the previous
section the strong coupling OPE has the form 
\ba
O^{I_1}(x_1)O^{I_2}(x_2)&=&\frac{\d^{I_1I_2}}{x_{12}^4}
+\frac{2^{3/2}}{N}C^{I_1I_2I}\frac{1}{x_{12}^2}[\,O^I\,]
+\frac{2^{7 /2}\pi^2}{3\lambda N}\frac{x_{12}^{\mu}}{x_{12}^2}
C^{I_1I_2}_{\J_{15}}[\,R^{\J_{15}}_{\mu}\,]\nonumber \\
&-&\frac{2\pi^2}{15\lambda N }\d^{I_1I_2}
\frac{x_{12}^{\mu}x_{12}^{\nu}}{x_{12}^2}[\,T_{\mu\nu}\,]
+\d^{I_1I_2}x^{\D_{1}^{(1)}}_{12}[\,O_{1}\,]\nonumber \\
&+&
C^{I_1I_2}_{{\cal J}_{20}}x^{\D_{20}^{(1)}}_{12}
[\,O^{{\cal J}_{20}}\,]
+C^{I_1I_2}_{{\cal J}_{105}}x^{\D_{105}^{(1)}}_{12}
[\,O^{{\cal J}_{105}}\,]
+
C^{I_1I_2}_{{\cal J}_{84}}x^{\D_{84}^{(1)}}_{12}
[\,O^{{\cal J}_{84}}\,]\nonumber \\
&+&C^{I_1I_2}_{{\cal J}_{15}}x^{\D_{15}^{(1)}}_{12}x^\mu_{12}
[\,O_\mu^{{\cal J}_{15}}\,]
+C^{I_1I_2}_{{\cal J}_{175}}x^{\D_{175}^{(1)}}_{12}x^\mu_{12}
[\,O_\mu^{{\cal J}_{175}}\,]+\ldots .
\label{strongope}
\ea
Here $R^{\J_{15}}_{\mu}$ is the R-symmetry current and $T_{\mu\nu}$ is
the stress tensor of $\N =4$ SYM$_4$. The structure constants of the
operators $O^I$, $R^{\J_{15}}_{\mu}$, $T_{\mu\nu}$ are found by
requiring that the above OPE 
reproduces the known 3-point functions of two CPOs with another CPO, 
the $R$-symmetry current and the stress tensor respectively, as the
latter were computed
in the supergravity approximation in \cite{LMRS,AF5}. 
The operator algebra structure constants of the double-trace
operators in (\ref{strongope}) are chosen to be 1, which means that 
their 2- and 3-point normalization constants are kept equal.
The anomalous
dimensions $\D_1, \D_{20},\ldots , \D_{175}$ of the double-trace
operators  will be determined in the next section by studying the 4-point
functions of the CPOs. 

Comparing (\ref{strongope}) with (\ref{freeope}), we see
that the structure of the strong coupling OPE is simpler than the
corresponding free field theory one. Instead of having an infinite number of
single-trace operators as in 
(\ref{freeope}), we find in (\ref{strongope}) only three
single-trace operators 
giving rise to the most singular terms. The 
coefficients in front of the $R$-symmetry current and the stress tensor are,
however, different from the ones in (\ref{freeope}). The reason is 
that the free field operators $J_\mu^{\J_{15}}$ and $T_{\mu\nu}^{fr}$
receiving 
contribution only from bosons may be represented as 
\bea
J_\mu^{\J_{15}}=\frac13 R_\mu^{\J_{15}}+  \frac23 {\cal K}_\mu^{\J_{15}};\quad
T_{\mu\nu}^{fr} = \frac15 T_{\mu\nu} + \frac{10}{35}{\cal K}_{\mu\nu}
+\frac{18}{35}\Xi_{\mu\nu},
\label{split}
\eea
where  ${\cal K}_\mu^{\J_{15}}$ and  ${\cal K}_{\mu\nu}$ are  vector and tensor
operators from the Konishi supermultiplet which has as  leading
component that scalar $\cal K$,  
while $\Xi_{\mu\nu}$ is the leading component of a new supersymmetry
multiplet.  
The splitting (\ref{split}) is explained by the fact that 
$T_{\mu\nu}$, ${\cal K}_{\mu\nu}$ and $\Xi_{\mu\nu}$ have pairwise vanishing 
two-point functions \cite{An,An1} and belong to different supersymmetry
multiplets. The operators in the Konishi supermultiplet as well as
$\Xi_{\mu\nu}$ are dual  
to string modes and therefore decouple in the strong coupling limit. 

A splitting analogous to (\ref{split}) may also occur for 
the free field theory double-trace operators. However, 
there is an important difference. If we assume that all operators
have free field theory 2-point normalization constants of order 1, then
the splitting has the following schematic form
$$O^{fr} = O^{gr} + \frac{1}{N}O^{str},$$
where a free field theory double-trace operator $O^{fr}$ 
is split into a sum of operators $O^{gr}$ dual to supergravity
multi-particle states, and operators $O^{str}$ dual to 
string states. As follows from the discussion 
in the previous section the coefficient
in front of $O^{str}$ has to be of order $1/N$, because otherwise one
would not reproduce the disconnected part of the  4-point function.
Such a splitting manifests itself in the $1/N^2$ corrections to 2- and 
3-point normalization constants of double-trace operators.
In what follows we will be mostly interested in double-trace operators
with  free-field dimensions 4 and 5. We will see that such a splitting
does occur for all the operators except the operators in the 
{\bf 20} and {\bf 105} irreps.

%%%%%%%%%%%%%%%%%%%%%%%%%%%%%%%%%%%%%%%%%%%%%%%%%%%%%%%%%%%%%%%%%%%%%%%%%%%%%
\section{Anomalous dimensions of double-trace operators}
\setcounter{equation}{0}
%%%%%%%%%%%%%%%%%%%%%%%%%%%%%%%%%%%%%%%%%%%%%%%%%%%%%%%%%%%%%%%%%%%%%%%%%%%%%%  
In this section we determine the anomalous dimensions of double-trace 
operators and the leading $1/N^2$ corrections to their 2- and 
3-point function normalization constants
$C_{D}(N)$. To this end, we study the asymptotic behavior of the 4-point
functions of CPOs in the direct channel $x_{12}^2,x_{34}^2 \to 0$.
Since we know all the 4-point functions, we do not need to consider 
the crossed channels. It is well-known that a conformally-invariant 4-point 
function is given as a general analytic function of two
variables, which are here conveniently chosen to be the ``biharmonic ratios''  
$$
u=\frac{x_{12}^2x_{34}^2}{x_{13}^2x_{24}^2}
\,,\,\,\,\,\,v=\frac{x_{12}^2x_{34}^2}{x_{14}^2x_{23}^2}.
$$
We also use in the following the variable $Y=1-\frac{v}{u}$.
The biharmonic ratios above and the variable $Y$ have the property that
$u,\,v,\,Y\,\rightarrow 0$ as 
$x_{12}^2,\,x_{34}^2\,\rightarrow 0$. 

To perform the computation we need to know  the
contributions of various quasi-primary operators and their descendents 
in the 4-point functions of CPOs, i.e. the conformal partial wave amplitudes of
quasi-primary operators. We restrict ourselves mainly to  
the contributions of scalar, vector and second rank symmetric
traceless tensor operators. Let the OPE of CPOs be of the form
\ba
O^{I_1}(x_1)O^{I_2}(x_2)&=&
C^{I_1I_2}_{{\cal J}}\biggl( \frac{C_{OOS}}{C_{S}}
\frac{1}{x^{4-\D_S}_{12}}[\,S^{{\cal J}}\,]
+\frac{C_{OOT}}{C_{T}}
\frac{x_{12}^\mu x_{12}^\nu}{x^{6-\D_T}_{12}}
[\,T^{{\cal J}}_{\mu\nu}\,]\nonumber \\
&~&~~~~~~+
\frac{C_{OOV}}{C_{V}}
\frac{x_{12}^\mu }{x^{5-\D_V}_{12}}
[\,V^{{\cal J}}_{\mu}\,]+\ldots \biggr) ,
\label{opestv}
\ea
where $\J$ denotes an index of an irreducible representation of 
the $R$-symmetry 
group $SO(6)$, $C^{I_1I_2}_{{\cal J}}$ are the 
Clebsh-Gordon coefficients  and
$\D_S,\,\D_T,\,\D_V$ are the conformal dimensions of the scalar, 
tensor and vector
operators  respectively. For any of the operators, $C_{\cal O}$ 
and $C_{OO{\cal O}}$ denote the normalization constant in the 2-point
function $\langle{\cal O}(x_1){\cal O}(x_2)\rangle$ and
the coupling constant in the three-point function 
$\langle O^I(x_1)O^J(x_2) {\cal O}(x_3)\rangle$, respectively. 
Then, one can show that the short-distance expansion of the conformal
partial amplitudes of the scalar S, tensor T and vector V operators
can be written as \cite{P}
\ba
&&\langle
  O^{I_1}(x_1)O^{I_2}(x_2)O^{I_3}(x_3)
  O^{I_4}(x_4)\rangle = \frac{C^{I_1I_2}_{{\cal J}}C^{I_3I_4}_{{\cal J}}}
{x_{12}^4x_{34}^4}
\nonumber \\
&&\hspace{1.5cm}\times
\Biggl[ 
\frac{C_{OOS}^2}{C_S}v^{\frac{\Delta_S}{2}} 
\Biggl( 1+\frac{\Delta_S}{4}Y
+\frac{\Delta_S^3}{16(\Delta_S -1)(\Delta_S+1)}v\left( 1
+\frac{\Delta_S+2}{4}Y\right) +\cdots\Biggl) 
\nonumber \\
&& \hspace{3cm}+\frac{C_{OOT}^2}{C_T}v^{\frac{\Delta_T}{2}-1}\Biggl(
\frac{1}{4}Y^2 -\frac{1}{\Delta_T}v-\frac{1}{\Delta_T}vY\cdots\Biggl)
\nonumber \\
&& \hspace{3.5cm}+\frac{C_{OOV}^2}{C_V}v^{\frac{\Delta_V -1}{2}}\Biggl(
\frac{1}{2}Y +\cdots\Biggl)\Biggl] .
\label{OPEa}
\ea
The formulas for the leading contributions of a rank-2 traceless 
symmetric tensor and a vector can be generalized to
the case of a rank-$l$ traceless symmetric tensor of dimension $\D_l$ and
one gets a leading term of the form
$$v^{\frac{\Delta_l -l}{2}}\,Y^l.$$
For this reason a term of the form $v^{\D/2}F(Y)$ in a 4-point function
contains, in principle,  the contributions not only from a scalar operator, but
also from any symmetric tensor operator of rank $l$ and 
conformal dimension $\D +l$.
Moreover, (\ref{OPEa}) shows that the anomalous dimensions are related
to terms of the type $v^{\frac{\D_S^{(0)}}{2}}\log v$ for scalar operators,
$v^{\frac{\D_V^{(0)}-1}{2}}\,Y\,\log v$ for vector operators and
$v^{\frac{\D_T^{(0)}-2}{2}}\,Y^2\,\log v$ for rank-2 tensor operators.

The 4-point functions of CPOs were computed in the supergravity 
approximation in \cite{AF7} and can be written
as follows
\ba
&&\langle O^{I_1}(x_1)O^{I_2}(x_2)O^{I_3}(x_3)O^{I_4}(x_4) \rangle
=\frac{\d^{I_1I_2}\d^{I_3I_4}}{x_{12}^4x_{34}^4}
+\frac{\d^{I_1I_3}\d^{I_2I_4}}{x_{13}^4x_{24}^4}
+\frac{\d^{I_1I_4}\d^{I_2I_3}}{x_{14}^4x_{23}^4}\cr
&&+\frac{8}{N^2\pi^2}\Biggl[-\frac{C^{-}_{I_1I_2I_3I_4}}
{x_{12}^2x_{34}^2}\Bigl( 2(x_{13}^2x_{24}^2
-x_{14}^2x_{23}^2)D_{2222}\nonumber \\
&& \hspace{4cm}-x_{24}^2D_{1212}-x_{13}^2D_{2121}+x_{14}^2D_{2112}
+x_{23}^2D_{1221} \Bigl)
\nonumber \\
&&+\d^{I_1I_2}\d^{I_3I_4}
\left( -\frac{1}{2x_{34}^2}D_{2211}+
\frac{(x_{13}^2x_{24}^2+ x_{14}^2x_{23}^2  
 -x_{12}^2x_{34}^2)}{x_{34}^2}
D_{3322}+ \frac{3}{2}D_{2222} \right)
\nonumber \\
&&+2
C^{+}_{I_1I_2I_3I_4}\left( \frac{1}{x_{34}^2}D_{2211}
+4x_{34}^2D_{2233}-3D_{2222} \right) +t+u\Biggr] , 
\label{4p}
\ea
where $C^{\pm}_{I_1I_2I_3I_4}=\frac12\left( C_{I_1I_2I_3I_4}
\pm C_{I_2I_1I_3I_4}\right)$ and $t$ and $u$ stand for the 
contributions of the $t$- and $u$-channels obtained by the 
interchange $1\leftrightarrow 4$
and $1\leftrightarrow 3$, respectively. The $D$-functions are defined as
\ba
&& \hspace{-.7cm}D_{\D_1\D_2\D_3\D_4}(x_1,x_2,x_3,x_4) =
\label{D1}\\
&&\hspace{-0.4cm} =\int\!\rmd^{d+1}\hat{x}\frac{x_0^{
    -d-1+\D_1+\D_2+\D_3+\D_4}}{[x_0^2+(x-x_1)^2]^{
    \D_1}[x_0^2+(x-x_2)^2]^{ \D_2} [x_0^2+(x-x_3)^2]^{ \D_3}
  [x_0^2+(x-x_4)^2]^{ \D_4}}.\nonumber
\ea
It is convenient to represent $D$-functions in the form
\ba
&&\hspace{-2cm}D_{\D_1\D_2\D_3\D_4}(x_1,x_2,x_3,x_4) =\nonumber \\
&& =\frac{\bar{D}_{\D_1\D_2\D_3\D_4}(v,Y)}
{(x_{12}^2)^{\frac{\D_1+\D_2-\D_3-\D_4}{2}}  (x_{13}^2)^{\frac{\D_1+\D_3-\D_2-\D_4}{2}}
(x_{23}^2)^{\frac{\D_2+\D_3+\D_4-\D_1}{2}} 
(x_{14}^2)^{\D_4}}.
\nonumber
\ea
As shown in Appendix C, a $\bar{D}$-function is given by a convergent 
series in $v$ and $Y$. 
In terms of the biharmonic ratios $u$ and $v$ the 4-point function 
acquires the form
\ba
&&\langle O^{I_1}(x_1)O^{I_2}(x_2)O^{I_3}(x_3)O^{I_4}(x_4)\rangle
=\frac{1}{x_{12}^4x_{34}^4}\biggl[
\d^{I_1I_2}\d^{I_3I_4}+u^2\d^{I_1I_3}\d^{I_2I_4}
+v^2\d^{I_1I_4}\d^{I_2I_3} 
\biggl]\cr
&&~~+\frac{8}{\pi^2 N^2}\frac{1}{x_{12}^4x_{34}^4}\Biggl\{
C_{I_1I_2I_3I_4}^-\biggl[ 
\bar{D}_{2222}\left(2 v- 2\frac{v^2}{u} +vu- v^2u-
\frac{v^3}{u}+ v^3\right)\nonumber \\     
&&~~~~~~~~~~~~~+\bar{D}_{1212}\left( \frac{2v^2}{u}-v^2+\frac{v^3}{u}\right) +
\bar{D}_{2112}\left( -2v-vu+v^2\right)\nonumber \\
&&~~~~~~~~~~~~~+\bar{D}_{2211}\left( vu-v^2\right) +
\bar{D}_{2323}\left( -\frac{4v^3}{u}\right) +
4v^2\bar{D}_{3223}\biggl]\nonumber \\
&&~~~~~+C_{I_1I_2I_3I_4}^+\biggl[
\bar{D}_{2222}\left( -12v^2- vu -
\frac{v^3}{u}+v^2u+ v^3\right)\label{4pointuv}\\     
&&~~~~~~~~~~~~~+\bar{D}_{1212}\left( v^2+\frac{v^3}{u}\right) +
\bar{D}_{2112}\left( vu+v^2\right)\nonumber \\
&&~~~~~~~~~~~~~+\bar{D}_{2211}\left( 2v-vu-v^2\right) +8v^2\bar{D}_{3322}+
 \frac{4v^3}{u}\bar{D}_{2323} + 4v^2\bar{D}_{3223}\biggr]\nonumber \\
&&~~~~~+C_{I_1I_3I_2I_4}\biggl[
\bar{D}_{2222}\left( -6v^2+vu
+\frac{v^3}{u}-v^2 u- v^3\right)\cr     
&&~~~~~~~~~~~~~+\bar{D}_{1212}\left( v^2-\frac{v^3}{u}\right) +
\bar{D}_{2112}\left( -vu+v^2\right)\nonumber \\
&&~~~~~~~~~~~~~+\bar{D}_{2211}\left( vu+v^2\right) +
 \frac{4v^3}{u}\bar{D}_{2323} + 4v^2\bar{D}_{3223}\biggr]\nonumber \\
&&~~~~~+\delta^{I_1I_2}\delta^{I_3I_4}\left[
\frac32 v^2\bar{D}_{2222}-\frac{1}{2}v\bar{D}_{2211}+
\bar{D}_{3322}(v+\frac{v^2}{u}-v^2)
\right] \nonumber \\
&&~~~~~+\delta^{I_1I_3}\delta^{I_2I_4}\left[
\frac{3}{2}v^2\bar{D}_{2222}-\frac{1}{2}v^2\bar{D}_{1212}+
\bar{D}_{2323}(v^2-\frac{v^3}{u}+v^3) \right]
\nonumber \\
&&~~~~~+\delta^{I_1I_4}\delta^{I_2I_3}\left[
\frac{3}{2}v^2\bar{D}_{2222}-\frac{1}{2}v^2\bar{D}_{2112}
+\bar{D}_{3223}(-v^2+\frac{v^3}{u}+v^3) \right] \Biggl\}.
\nonumber
\ea
This 4-point function is given as a sum of contributions from 
quasi-primary operators transforming in the six irreducible 
representations (\ref{irreps}) of $SO(6)$. It is clear that to obtain a 
contribution of operators belonging to a $D$-dimensional irrep
one should multiply the 4-point function by a $SO(6)$ tensor 
$C^{I_1I_2}_{{\cal J}_D}C^{I_3I_4}_{{\cal J}_D}$ which is a projector
onto the irrep. 
  
In what follows it will be sometimes useful to compare the short-distance expansion
of the 4-point function (\ref{4pointuv}) with the one of 
the free field 4-point function (\ref{free4}), which in
terms of the biharmonic ratios takes the form
\ba
&&\langle O^{I_1}(x_1)O^{I_2}(x_2)O^{I_3}(x_3)O^{I_4}(x_4)\rangle _{fr}=
\frac{1}{x_{12}^4x_{34}^4}\biggl[
\d^{I_1I_2}\d^{I_3I_4}+u^2\d^{I_1I_3}\d^{I_2I_4}
+v^2\d^{I_1I_4}\d^{I_2I_3} \cr
&&~~~~~~~~~~~~~+\frac{4}{N^2}\left( (u+v)C^{+}_{I_1I_2I_3I_4}
+(v-u)C^{-}_{I_1I_2I_3I_4}+uv C_{I_1I_3I_2I_4}
\right)\biggl].
\label{free4uv}
\ea
%%%%%%%%%%%%%%%%%%%%%%%%%%%%%%%%%%%%%%%%%%%%%%%%%%%%%%%%%%%%%%%%%%%%%%%%%%%%
\subsection{Projection on the singlet}
%%%%%%%%%%%%%%%%%%%%%%%%%%%%%%%%%%%%%%%%%%%%%%%%%%%%%%%%%%%%%%%%%%%%%%%%%%%%%
First we project the 4-point function on the singlet part that 
amounts to applying to it 
$\frac{1}{400}\delta^{I_1I_2}\delta^{I_3I_4}$. From the strong coupling OPE
(\ref{strongope}) we expect to find the stress tensor contribution 
and a contribution of the double-trace  
scalar operator $O_1$ of approximate dimension 4. 

The result for the connected part is 
\ba
&&\left. \langle O^{I_1}(x_1)O^{I_2}(x_2)O^{I_3}(x_3)O^{I_4}(x_4)\rangle
\right|_{{\bf 1}}
=\frac{8}{20\pi^2 N^2}\frac{\d^{I_1I_2}\d^{I_3I_4}}{x_{12}^4x_{34}^4}
\biggl[ \nonumber \\ 
&&
\bar{D}_{2222}\left( -9v^2-\frac{3 v^3}{u} -3 vu+3 v^2u
+3v^3\right)\nonumber \\     
~~~~~~~~&&+\bar{D}_{1212}\left( \frac{19}{6}v^2+3\frac{v^3}{u}\right) +
\bar{D}_{2112}\left( 3vu+\frac{19}{6}v^2\right)\cr
~~~~~~~~&&+\bar{D}_{2211}\left( -\frac{10}{3}v-3vu-3v^2\right) +
\bar{D}_{3322}\left( 20\frac{v^2}{u}+20v+\frac{20}{3}v^2
\right)\nonumber \\
~~~~~~~~&&+
\bar{D}_{2323}\left( v^2+ \frac{41v^3}{3u}+v^3\right)
+\bar{D}_{3223}\left( \frac{41}{3}v^2+\frac{v^3}{u}+v^3\right)
\biggr]
\nonumber
\ea
Using the formulas for the $\bar{D}$-functions from
the Appendix C, we can find that the most singular terms 
of the $v$-expansion are
\ba 
\left.\langle O^{I_1}(x_1)O^{I_2}(x_2)O^{I_3}(x_3)O^{I_4}(x_4)\rangle
\right|_{{\bf 1}} = 
\frac{\d^{I_1I_2}\d^{I_3I_4}}{N^2x_{12}^4x_{34}^4}\biggl[
vF_1(Y)+v^2F_2(Y)
+v^2\log v\,G_2(Y)\biggr],
\label{4p1}
\ea
where
\ba
F_1(Y)&=&\frac{4Y^2-8Y}{Y^3}
+\frac{4(-6+6Y-Y^2)\log (1-Y)}{3Y^3}\, ,\nonumber \\
F_2(Y)&=&\frac{ -1680+3360Y-2108Y^2+428Y^3
-21Y^4}{15(1-Y)Y^4}\, , \nonumber \\
&-&\frac{4\left( 1140-1890Y+962Y^2-151Y^3
+5Y^4\right)}{15Y^5}\log (1-Y)\nonumber \\
&+&
\frac{16(Y-2)(6-6Y+Y^2)}{Y^5}\Li (Y) \, ,
\nonumber \\
G_2(Y)&=&\frac{4(6-6Y+Y^2)}{3Y^4}
\left(\frac{12-12Y+Y^2}{Y-1}
+\frac{6\left( Y-2\right)
\log (1-Y)}{Y}
\right)\, .\nonumber
\ea
Expanding the functions in powers of $Y$ we then obtain
\ba
&&\left.\langle O^{I_1}(x_1)O^{I_2}(x_2)O^{I_3}(x_3)O^{I_4}(x_4)\rangle
\right|_{{\bf 1}} = 
\frac{1}{N^2}\frac{\d^{I_1I_2}\d^{I_3I_4}}{x_{12}^4x_{34}^4}
\biggl[ \frac{2}{45}vY^2 +
v^2\left( \frac{47}{225} -\frac{4}{5} \log v\right)\nonumber \\
&&\hspace{10cm}
-\frac{43}{225}v^2Y\biggr].
\label{4p11}
\ea
Comparing this asymptotics with (\ref{OPEa}), 
we see that the contribution from a scalar field of dimension 2 
is absent, as it should be, since the Konishi field acquires large anomalous
dimension and decouples in the strong coupling limit.  
We also get the relation:
\ba
\frac{C_{OOT}^2}{4C_T}=\frac{2}{45N^2}.\nonumber
\ea
Since for $C_{OOT}$ one has $C_{OOT}=\frac{4}{3\pi^2}\frac{\lambda}{N}$ 
\footnote{This value of the coupling constant is fixed by a conformal
Ward identity \cite{OP}, the same value was also obtained in the
supergravity approximation in \cite{AF5}. } 
one finds at strong coupling 
\ba
C_T=\frac{10\lambda^2}{\pi^4},\nonumber
\ea
which represents the normalization of the complete stress tensor of
the $\N =4$ SYM$_4$ \cite{LT}.  

As it was discussed above, a term of the form $vF(Y)$ contains, in general,
contributions from all traceless symmetric tensor operators
of rank $l$ and dimension $2+l$. However, comparing $F_1(Y)$ in 
(\ref{4p1}) with the corresponding term in the conformal partial wave 
amplitude of the stress tensor (\ref{cast}) we see that they 
coincide. Thus, the strong coupling OPE does not contain
single-trace rank-$l$ traceless symmetric tensors with dimension 
$2+l$ in its singlet part. Nevertheless, it may in principle contain tensors
of dimension $4+l$ or higher. However, as it was shown in
section 3 a possible single-trace scalar operator of 
dimension 4 vanishes. Thus the only scalar operator of 
approximate dimension 4 
is the double-trace operator $O_{{\bf 1}}$.\footnote{The free field
theory operator $O^{fr}_{\bf 1}$ probably splits into a linear
combination  of 
 $O_{\bf 1}$ and an operator $O^{str}_{\bf 1}$  
dual to a string mode. However, 
the coefficient in front of $O^{str}_{\bf 1}$ is of order $1/N$, 
and even if
the latter operator does not decouple in the strong coupling limit
it cannot contribute to $\log$-dependent terms in 4-point functions.
In the following, when discussing double-trace operators in
other irreps we simply assume that operators such as  $O^{str}_{\bf
  1}$ above do
decouple, making at the same time a consistency check to confirm  
our  assumption.}

The formula (\ref{4p11}) also allows us to determine the anomalous
dimension of $O_{{\bf 1}}$.
Assuming the existence at strong coupling of a scalar 
field with dimension $\Delta=\Delta^{(0)}+\Delta^{(1)}$, where 
$\Delta^{(0)}=4$ and $\Delta^{(1)}$ is the anomalous 
dimension, we find that 
$$
v^{\frac{\Delta}{2}}=v^{2}+\frac{1}{2}\Delta^{(1)}v^{2}\log v+...
$$
Since there is only one operator of approximate dimension 4,  
we do not face the problem  of operator mixing and 
from (\ref{OPEa}) we get 
$$
\frac{1}{2}\frac{C_{OOO_{\bf 1}}^2}{C_{O_{\bf 1}}}\Delta^{(1)} 
=-\frac{4}{5N^2}.
$$
Since $\Delta^{(1)}$ is of order $1/N^2$ we use for 
$\frac{C_{OOO_{\bf 1}}^2}{C_{O_{\bf 1}}}$
the $O(1)$ result which is $1/10$. In this way we obtain 
\ba
\Delta^{(1)}=-\frac{16}{N^2},
\ea 
for the anomalous dimension of $O_{\bf 1}$. 
This coincides with the anomalous dimension of
the operator $B$ considered in \cite{HMMR}, as it should be,
since $B$ is a descendent operator of $O_{{\bf 1}}$. 

We can also find the leading $1/N^2$ correction
to the 2- and 3-point normalization constant $C_{O_{{\bf 1}}}$. 
Writing as
$$C_{O_{{\bf 1}}}=\frac{1}{10}\left( 1
+\frac{1}{N^2}C_{O_{{\bf 1}}}^{(1)}\right),$$
and taking into account that 
$C_{O_{{\bf 1}}}=C_{OOO_{{\bf 1}}}$, 
we find from the term of order $v^2$
$$
C_{O_{{\bf 1}}}^{(1)}=\frac{38}{15}.
$$
Finally, we can make a consistency check of our computation.
Namely, since we know corrections to the conformal dimension, 
$\Delta^{(1)} =-16/N^2$
and to the 
structure constant we can compute the term of order $v^2Y$ by 
using (\ref{OPEa}), in order to compare it with the corresponding
value obtained from 
our 4-point  
function. Taking into account the contribution of the stress tensor 
we get from (\ref{OPEa}) and from the expansion of 
our 4-point function the same number $-\frac{43}{225}$. This also
confirms that there is only one operator of 
approximate dimension 4 in the strong coupling OPE, and that 
the operator $O^{str}_{\bf 1}$ decouples in the strong coupling limit.

We  can also compute the 2-point normalization constant in free field theory
by using (\ref{prop}) and the definition of the operator. A simple 
calculation gives
$$C_{O_{{\bf 1}}}=\frac{1}{10}\left( 1
+\frac{2}{3N^2}\right).$$
Thus, not only the conformal dimension but also the 2- and 3-point normalization
constants get $\frac{1}{N^2}$ corrections in the strong coupling limit.

%%%%%%%%%%%%%%%%%%%%%%%%%%%%%%%%%%%%%%%%%%%%%%%%%%%%%%%%%%%%%%%%%%%%%%%%%
\subsection{Projection on {\bf 20}}
%%%%%%%%%%%%%%%%%%%%%%%%%%%%%%%%%%%%%%%%%%%%%%%%%%%%%%%%%%%%%%%%%%%%%%%%%
According to (\ref{OPEa}), to obtain the contribution
of the operators transforming in a D-dimensional irrep, we should
multiply the 4-point function by the projector onto the representation
\ba
(P_{{\bf D}})_{I_1I_2I_3I_4}=\frac{1}{\nu_D}C^{I_1I_2}_{\J_{D}}
C^{I_3I_4}_{\J_{D}},
\label{projD}
\ea
where
$$\nu_D = \sum_{I_i}\,C^{I_1I_2}_{\J_{D}}C^{I_3I_4}_{\J_{D}}\,
C^{I_1I_2}_{\J'_{D}}C^{I_3I_4}_{\J'_{D}},
$$
is the dimension of the irrep so that  $P_{{\bf D}}^2=\frac{1}{\nu_D}$.

The projector on the ${\bf 20}$ can be easily found by taking into account
that the Clebsh-Gordon coefficient $C^{I_1I_2}_{\J_{20}}$ is proportional
to the $SO(6)$ tensor $C^{I_1I_2I_3}$. Then, one can show that
\ba
(P_{{\bf 20}})_{I_1I_2I_3I_4}=
\frac{3}{100}\left( C^{+}_{I_1I_2I_3I_4}-
\frac{1}{6}\d_{I_1I_2}\d_{I_3I_4}\right) .
\label{proj20}
\ea
Using the Table 1 from Appendix D for the contractions
of the projector with the $SO(6)$ tensors appearing in the 4-point function,
we find the contribution of the operators in the ${\bf 20}$ to 
the connected part of the 4-point function
\ba
&&\left. \langle O^{I_1}(x_1)O^{I_2}(x_2)O^{I_3}(x_3)O^{I_4}(x_4)\rangle
\right|_{{\bf 20}}
=\frac{8}{\pi^2 N^2}\frac{C^{I_1I_2}_{\J_{20}}C^{I_3I_4}_{\J_{20}}}
{x_{12}^4x_{34}^4}
\biggl[ \nonumber \\ 
&&~~~~~
\bar{D}_{2222}\left( -18v^2-\frac{3}{2} vu-
\frac{3v^3}{2u}+\frac{3}{2} v^2u+\frac{3}{2} v^3\right)\nonumber \\     
&&~~~~+\bar{D}_{1212}\left(\frac{4}{3}v^2+\frac{3v^3}{2u}\right) +
\bar{D}_{2112}\left(\frac{3}{2}vu+\frac{4}{3}v^2\right)
+\bar{D}_{2211}\left( \frac{10}{3}v-\frac{3}{2}vu-\frac{3}{2}v^2\right) \nonumber \\
&&~~~~+\frac{40}{3}v^2\bar{D}_{3322}+
\bar{D}_{2323}\left( v^2+\frac{19v^3}{3u}+v^3 \right)
+\bar{D}_{3223}\left(\frac{19}{3}v^2+\frac{v^3}{u}+v^3 \right) \biggr].
\label{4p20}
\ea
 Expanding the $\bar{D}$-functions in powers of $v$, we obtain 
\ba 
&&\left.\langle O^{I_1}(x_1)O^{I_2}(x_2)O^{I_3}(x_3)O^{I_4}(x_4)\rangle
\right|_{{\bf 20}} =
\frac{1}{N^2}\frac{C^{I_1I_2}_{\J_{20}}C^{I_3I_4}_{\J_{20}}}
{x_{12}^4x_{34}^4}\biggl[\nonumber \\
&&~~~~~~~~~~~~~
vF_1(Y)+v^2F_2(Y)
+v^2\log v\,G_2(Y)\biggr],
\label{4p200}
\ea
where
\ba
F_1(Y)&=&-\frac{40\log (1-Y)}{3Y},\nonumber \\
F_2(Y)&=&-\frac{ 8\left( 65-65Y+6Y^2\right)}{3(1-Y)Y^2} 
-\frac{20\left( 74-49Y+2Y^2\right)}{3Y^3}\log (1-Y)\nonumber \\
&+&
\frac{160(Y-2)}{Y^3}\Li (Y),
\nonumber \\
G_2(Y)&=&\frac{40}{3Y^2}
\left(\frac{12-12Y+Y^2}{Y-1}
+\frac{6\left( Y-2\right)
\log (1-Y)}{Y}
\right).\nonumber
\ea
Expanding the above functions in powers of $Y$ we finally obtain
\ba
\left.\langle O^{I_1}(x_1)O^{I_2}(x_2)O^{I_3}(x_3)O^{I_4}(x_4)
\rangle\right|_{{\bf 20}} &=& 
\frac{1}{N^2}\frac{C^{I_1I_2}_{\J_{20}}C^{I_3I_4}_{\J_{20}}}
{x_{12}^4x_{34}^4}\biggl[ \frac{40}{3}v +
\frac{26}{9}v^2\left( 1+Y\right) \nonumber \\
&&\hspace{3cm}
 -\frac{4}{3} v^2Y^2\log v\biggr].
\label{4p201}
\ea
The analysis of the results obtained follows the one in the 
previous subsection. Firstly, comparing $F_1(Y)$ in (\ref{4p200})
with the corresponding term of the conformal partial amplitude
of a scalar operator of dimension 2 (\ref{CPWAs}), we see that they
coincide.\footnote{Recall that for the lowest weight CPOs 
one has $\frac{C^2_{OOO}}{C_O}=\frac{40}{3N^2}$.} 
Therefore, all single-trace rank-$l$ traceless tensors of
dimension $2+l$ transforming in the ${\bf 20}$
 are absent in the OPE. Then, the only scalar operator of 
approximate dimension 4 is the double-trace operator $O_{\bf 20}$.
Moreover, we see that $\log v$-dependent terms appear
starting from the term $v^2Y^2\log v$. Thus we conclude from (\ref{OPEa})
that the double-trace operator $O_{\bf 20}$ has protected conformal
dimension. It is worth noting that the non-renormalization 
of the conformal dimension of this operator is not related to
the shortening condition discussed in \cite{AFer} and is a 
prediction of the AdS/CFT correspondence. 
The first operators which acquire anomalous dimensions are scalar
and tensor operators of approximate dimension 6.  
  
The first $1/N^2$ correction
to the 2- and 3-point normalization constant $C_{O_{{\bf 20}}}$ 
can also be easily found. 
Writing the constant as
$$C_{O_{{\bf 20}}}=2\left( 1
+\frac{1}{N^2}C_{O_{{\bf 20}}}^{(1)}\right),$$
and taking into account the contribution of the single-trace operator
$O^I$ and that 
$C_{O_{{\bf 20}}}=C_{OOO_{{\bf 20}}}$, 
we find from the term of order $v^2$
$$
C_{O_{{\bf 20}}}^{(1)}=\frac{1}{3}\, .
$$
The 2-point normalization constant can be also computed 
in free field theory by using (\ref{prop}) and the 
definition of the operator (\ref{doubtr}) and appears to
coincide with the value obtained in the strong coupling limit 
$$C_{O_{{\bf 20}}}=2\left( 1
+\frac{1}{3N^2}\right).$$
Thus, both the conformal dimension and the 2-point function normalization
constant of the double-trace operator in the ${\bf 20}$ are 
non-renormalized in the strong coupling limit. This also shows that
in this case there is no splitting, and the free field theory double-trace
operator coincides with $O_{\bf 20}$.
%%%%%%%%%%%%%%%%%%%%%%%%%%%%%%%%%%%%%%%%%%%%%%%%%%%%%%%%%%%%%%%%%%%%%%%%%
\subsection{Projection on {\bf 105}}
%%%%%%%%%%%%%%%%%%%%%%%%%%%%%%%%%%%%%%%%%%%%%%%%%%%%%%%%%%%%%%%%%%%%%%%%%
The free field theory OPE (\ref{freeope}) and the strong coupling
OPE (\ref{strongope}) do not contain single-trace operators 
transforming in the {\bf 105} irrep. Thus, only double-trace
operators contribute to this part of the 4-point function. 
The corresponding connected contribution can be easily found using
the Table 1 from Appendix D and is given by
\ba
&&\left. \langle O^{I_1}(x_1)O^{I_2}(x_2)O^{I_3}(x_3)O^{I_4}(x_4)\rangle
\right|_{{\bf 105}}
=\frac{8}{\pi^2 N^2}\frac{C^{I_1I_2}_{\J_{105}}C^{I_3I_4}_{\J_{105}}}
{x_{12}^4x_{34}^4}
\biggl[ \nonumber \\ 
&&~~~~~~~~\bar{D}_{2222}\left( -3v^2+vu
+\frac{v^3}{u}- v^2u-v^3\right)\nonumber \\     
&&~~~~~~+\bar{D}_{1212}\left(\frac{1}{2} v^2-\frac{v^3}{u}\right) +
\bar{D}_{2112}\left( -vu+\frac{1}{2}v^2\right)
+\bar{D}_{2211}\left( vu+v^2\right) \nonumber \\
&&~~~~~~+
\bar{D}_{2323}\left( \frac{3v^3}{u}+v^3+v^2\right)
+ \bar{D}_{3223}\left(3v^2+v^3+\frac{v^3}{3}\right)
\biggr].
\label{4p105}
\ea
 Expanding the $\bar{D}$-functions in powers of $v$, we obtain 
\ba 
\left.\langle O^{I_1}(x_1)O^{I_2}(x_2)O^{I_3}(x_3)O^{I_4}(x_4)\rangle
\right|_{{\bf 105}} &=& 
\frac{1}{N^2}\frac{C^{I_1I_2}_{\J_{105}}C^{I_3I_4}_{\J_{105}}}
{x_{12}^4x_{34}^4}\biggl[
v^2F_2(Y)+v^3F_3(Y)\nonumber \\
&+&v^4F_4(Y)
+v^4\log v\,G_4(Y)\biggr],
\label{4p1050}
\ea
where
\ba
F_2(Y)&=&\frac{4}{1-Y},\nonumber \\
F_3(Y)&=&\frac{ 4\left( Y-2\right)}{(1-Y)Y^2} 
-\frac{8}{Y^3}\log (1-Y),\nonumber \\
F_4(Y)&=&-\frac{ 4\left( 28-28Y+3Y^2\right)}{(1-Y)Y^4} 
-\frac{8\left( 38-25Y+Y^2\right)}{Y^5}\log (1-Y)\nonumber \\
&+&
\frac{96(Y-2)}{Y^5}\Li (Y),
\nonumber \\
G_4(Y)&=&\frac{8}{Y^4}
\left(\frac{12-12Y+Y^2}{Y-1}
+\frac{6\left( Y-2\right)
\log (1-Y)}{Y}
\right).\nonumber
\ea
Since only double-trace  operators contribute, it is useful to
compare (\ref{4p1050}) with the corresponding part of the free
field theory 4-point function (\ref{free4uv})
\ba 
&&\left.\langle O^{I_1}(x_1)O^{I_2}(x_2)O^{I_3}(x_3)O^{I_4}(x_4)\rangle_{fr}
\right|_{{\bf 105}} = 
\frac{C^{I_1I_2}_{\J_{105}}C^{I_3I_4}_{\J_{105}}}
{x_{12}^4x_{34}^4}\biggl[\nonumber \\
&&~~~~~v^2\left( 1+\frac{1}{(1-Y)^2}\right) + \frac{v^2}{N^2}\frac{4}{1-Y}
\biggr].
\label{4p105free}
\ea
The first term on the r.h.s. of this equation shows the disconnected part 
of the free field theory 4-point function. Comparing the term of order
$1/N^2$ in (\ref{4p105free}) with the term $v^2F_2(Y)$ in (\ref{4p1050}),
we see that they coincide. This means that the conformal
dimensions and the leading  corrections in $1/N^2$ to 2- and 3-point functions
normalization constants of any
symmetric traceless rank-$2k$ tensor operator of dimension 
$4+2k$ transforming in the ${\bf 105}$ coincide with the ones
computed in free field theory. Thus, all these operators are non-renormalized
in the strong coupling limit. The first correction to the 2- and 3-point 
functions normalization constant of the double-trace operator $O_{\bf 105}$
can be easily found from (\ref{4p105free}) and is given by
$$C_{O_{{\bf 105}}}=2\left( 1
+\frac{2}{N^2}\right).$$
The non-renormalization of the double-trace operator $O_{\bf 105}$ 
follows from the shortening conditions derived in \cite{AFer,AFer2},
and was also checked in perturbation theory at small YM coupling in
\cite{AFer,BKRS1,Sk,BKRS2}.

The expansion (\ref{4p1050}) also shows that the first $\log v$-term
appears at order $v^4$. Therefore, all 
symmetric traceless rank-$2k$ tensor operators of dimension 
$6+2k$ transforming in the ${\bf 105}$ have protected conformal
dimensions. Note, however, that the normalization constants of their 
2- and 3-point functions certainly 
receive corrections at strong coupling,
which are encoded in the function $F_3(Y)$. The vanishing of
anomalous dimensions of these tensor operators does not seem to follow
from any known non-renormalization theorem. These results also
demonstrate that the free field theory 
symmetric traceless rank-$2k$ tensor operators of dimension 
$4+2k$ do not split, while the ones with dimension $6+2k$ 
do. 
  
Since $G_4(Y)=-\frac{4}{5}+...$ 
the first double-trace operator in  the ${\bf 105}$ 
which acquires anomalous dimension is the scalar operator with
approximate dimension 8.  
%%%%%%%%%%%%%%%%%%%%%%%%%%%%%%%%%%%%%%%%%%%%%%%%%%%%%%%%%%%%%%%%%%%%%%%%%
\subsection{Projection on ${\bf 84}$}
%%%%%%%%%%%%%%%%%%%%%%%%%%%%%%%%%%%%%%%%%%%%%%%%%%%%%%%%%%%%%%%%%%%%%%%%%
Just as it was the case for the operators in the {\bf 105}, only double-trace
operators transforming in the {\bf 84} irrep can contribute to this 
part of the 4-point function. 
The corresponding connected contribution is again found by using
the Table 1 from Appendix D:
\ba
&&\left. \langle O^{I_1}(x_1)O^{I_2}(x_2)O^{I_3}(x_3)O^{I_4}(x_4)\rangle
\right|_{{\bf 84}}
=\frac{8}{\pi^2 N^2}\frac{C^{I_1I_2}_{\J_{84}}C^{I_3I_4}_{\J_{84}}}
{x_{12}^4x_{34}^4}
\biggl[ \nonumber \\ 
&&~~~~~~~\bar{D}_{2222}\left(6v^2-\frac{uv}{2}+\frac{uv^2}{2}
+\frac{v^3}{2}-\frac{v^3}{2u}\right)\nonumber \\     
&&~~~~~~~+\bar{D}_{1212}\left( -v^2+\frac{v^3}{2u}\right) +
\bar{D}_{2112}\left(\frac{uv}{2}-v^2\right)
-\bar{D}_{2211}\left( \frac{uv}{2}+\frac{v^2}{2}\right) \nonumber \\
&&~~~~~~~+
\bar{D}_{2323}\left(v^2+v^3-\frac{3v^3}{u}\right)
+\bar{D}_{3223}\left(-3v^2+v^3+\frac{v^3}{u}\right)
\biggr]\, .
\label{4p84}
\ea
 Expanding the $\bar{D}$-functions in powers of $v$, we obtain 
\ba 
&&\left.\langle O^{I_1}(x_1)O^{I_2}(x_2)O^{I_3}(x_3)O^{I_4}(x_4)\rangle
\right|_{{\bf 84}} =
\frac{1}{N^2}\frac{C^{I_1I_2}_{\J_{84}}C^{I_3I_4}_{\J_{84}}}
{x_{12}^4x_{34}^4}\biggl[v^2F_2(Y)+v^3F_3(Y)\nonumber \\
&&\hspace{9.5cm}
+v^3\log v\,G_3(Y)\biggr],
\label{4p840}
\ea
where
\ba
F_2(Y)&=&-\frac{ 8\left( 3-3Y+Y^2\right)}{(1-Y)Y^2} 
+\frac{12(Y-2)}{Y^3}\log (1-Y)\, ,\nonumber \\
F_3(Y)&=&\frac{ 8(Y-2)\left( 21-21Y+2Y^2\right)}{(1-Y)Y^4} 
+\frac{4\left( -228+264Y-80Y^2+3Y^3\right)}{Y^5}\log (1-Y)\nonumber \\
&-&
\frac{144(Y-2)^2}{Y^5}\Li (Y)\, ,
\nonumber \\
G_3(Y)&=&-\frac{12(Y-2)}{Y^4}
\left(\frac{12-12Y+Y^2}{Y-1}
+\frac{6\left( Y-2\right)
\log (1-Y)}{Y}
\right).\nonumber
\ea
Since the first $\log v$-term
appears at order $v^3$, all 
symmetric traceless rank-$2k$ tensor operators of dimension 
$4+2k$ transforming in the ${\bf 84}$ have protected conformal
dimensions. The first double-trace operator in  the ${\bf 84}$ 
which acquires an anomalous dimension is the scalar operator with 
approximate dimension 6. However, contrary to the case of the {\bf 105}
irrep, the leading  $1/N^2$ corrections to the normalization constants of 
the 2- and 3-point functions of these operators differ from their 
free field theory values. To see this we  
compare (\ref{4p840}) with the corresponding part of the free
field theory 4-point function (\ref{free4uv})
\ba 
&&\left.\langle O^{I_1}(x_1)O^{I_2}(x_2)O^{I_3}(x_3)O^{I_4}(x_4)\rangle_{fr}
\right|_{{\bf 84}} = 
\frac{C^{I_1I_2}_{\J_{84}}C^{I_3I_4}_{\J_{84}}}
{x_{12}^4x_{34}^4}\biggl[v^2\left( 1+\frac{1}{(1-Y)^2}\right)
\nonumber \\
&&\hspace{10cm}- \frac{v^2}{N^2}\frac{2}{1-Y}
\biggr].
\label{4p84free}
\ea
Expanding (\ref{4p840}) and  (\ref{4p84free}) in powers of $Y$, we
obtain the  normalization constants of 
2- and 3-point functions of the operator
$O_{\bf 84}$ at strong coupling and in free field theory
correspondingly as
\ba
C_{O_{{\bf 84}}}^{str}&=&2\left( 1
-\frac{3}{N^2}\right),\nonumber \\
C_{O_{{\bf 84}}}^{fr}&=&2\left( 1
-\frac{1}{N^2}\right).
\nonumber
\ea
The vanishing of the anomalous dimensions of the double-trace operator 
$O_{\bf 84}$ follows from the shortening conditions discussed in 
\cite{AFer,AFer2} and was also shown in perturbation theory 
at small YM coupling in \cite{Sk,BKRS2}. 
The difference between $C_{O_{\bf 84}}^{str}$ and 
$C_{O_{\bf 84}}^{fr}$ again may find a natural explanation in the fact 
that the corresponding free field theory operator 
undergoes a linear splitting on $O_{\bf 84}$ and ${\cal K}_{\bf 84}$,
where $O_{\bf 84}$ has protected both its dimension and the
normalization constants of the 2- and 3-point functions, while 
the operator ${\cal K}_{\bf 84}$ belongs to the Konishi 
multiplet \cite{BKRS2} and, therefore, decouples at strong coupling.
%%%%%%%%%%%%%%%%%%%%%%%%%%%%%%%%%%%%%%%%%%%%%%%%%%%%%%%%%%%%%%%%%%%%%%%%%
\subsection{Projection on ${\bf 15}$}
%%%%%%%%%%%%%%%%%%%%%%%%%%%%%%%%%%%%%%%%%%%%%%%%%%%%%%%%%%%%%%%%%%%%%%%%%
By using the projector $(P_{{\bf 15}})_{I_1I_2I_3I_4}$ 
constructed in the 
Appendix D and the results of Table 1 we find 
the following contribution
of the operators in ${\bf 15}$ to the connected 
part of the 4-point function
\ba
&&\left. \langle O^{I_1}(x_1)O^{I_2}(x_2)O^{I_3}(x_3)O^{I_4}(x_4)\rangle
\right|_{{\bf 15}}
=\frac{8}{\pi^2 N^2}\frac{C^{I_1I_2}_{\J_{15}}C^{I_3I_4}_{\J_{15}}}{x_{12}^4x_{34}^4}
\biggl[ \nonumber \\
&&\bar{D}_{2222}\left(4v+2uv-\frac{4v^2}{u}-2uv^2+2v^3
-\frac{2v^3}{u}\right)\nonumber \\     
&&+\bar{D}_{1212}\left( -\frac{3v^2}{2}+\frac{4v^2}{u}
+\frac{2v^3}{u}\right) +
\bar{D}_{2112}\left(-4v -2uv+\frac{3v^2}{2}\right)\nonumber \\
&&+\bar{D}_{2211}\left( 2vu-2v^2\right) +
\left(-v^2-v^3-\frac{7v^3}{u}\right)\bar{D}_{2323} 
+\left(7v^2+v^3+\frac{v^3}{2u}\right)\bar{D}_{3223} 
\biggr].
\nonumber
\ea
Expansion of the $D$-functions in powers of $v$ 
produces now the following
expression for leading terms 
\ba
\nonumber
\left. \langle O^{I_1}(x_1)O^{I_2}(x_2)O^{I_3}(x_3)O^{I_4}(x_4)\rangle
\right|_{{\bf 15}}
&=&\frac{1}{N^2}\frac{C^{I_1I_2}_{\J_{15}}C^{I_3I_4}_{\J_{15}}}
{x_{12}^4x_{34}^4}
\biggl[ vF_1(Y)+v^2 F_2(Y)\nonumber \\
&&\hspace{1cm}+v^2\log v G_2(Y)\biggl].
\ea
Here the functions $F_1,F_2$ and $G_2$ are given by 
\ba
F_1(Y)&=&\frac{16}{Y^2}\left(-2Y+(Y-2)\log(1-Y)\right), \nonumber \\
F_2(Y)&=&-\frac{4(Y-2)(56-56Y+5Y^2)}{(Y-1)Y^3}+
\frac{8(-152+176Y-53Y^2+2Y^3)}{Y^4}\log(1-Y) \nonumber \\
&-&\frac{192(Y-2)^2}{Y^4}\Li(Y), \nonumber \\
\nonumber
G_2(Y)&=&-\frac{16(Y-2)}{Y^3}\left(\frac{12-12Y+Y^2}{Y-1}
+\frac{6(Y-2)\log(1-Y)}{Y}\right).
\ea
Expansion in powers of $Y$ produces the following leading terms  
\ba
\label{4pt15}
\left. \langle O^{I_1}(x_1)O^{I_2}(x_2)O^{I_3}(x_3)O^{I_4}(x_4)\rangle
\right|_{{\bf 15}}
=\frac{C^{I_1I_2}_{\J_{15}}
C^{I_3I_4}_{\J_{15}}}{N^2x_{12}^4x_{34}^4}
\biggl[\frac{8}{3}vY+\frac{12}{25}v^2Y-\frac{16}{5}v^2Y\log v\biggl].
\ea
The absence of the terms $vY\log v$ shows that the vector operator 
of the dimension 3, which is the $R$-symmetry 
current $R_{\mu}^{{\cal J}_{15}}$, 
has protected conformal dimension. According to the 
discussion above, the function $F_1(Y)$ may 
receive contributions 
from single-trace rank $2k+1$ traceless 
tensors of dimension
$2k+3$ transforming in ${\bf 15}$, which is what  
indeed happens in the free field 
theory limit. However, comparing the function $F_1(Y)$
with the relevant part of the conformal partial amplitude
of the conserved vector current of dimension 3 (\ref{cavc})
one concludes that they coincide, therefore, the corresponding 
tensors are absent in the strong-coupling OPE. Next, comparing 
(\ref{4pt15})   
with eq.(\ref{OPEa}) we read off the value of the ratio
$$
\frac{C_{OOR}^2}{2C_R}=\frac{8}{3N^2}.
$$  
Since the value of $C_{OOR}$ is fixed by the conformal Ward identity
to be $C_{OOR}=\frac{2^{1/2}}{\pi^2}\frac{\lambda}{N}$ one finds 
$$
C_{R}=\frac{3\lambda^2}{8\pi^4}
$$
which corresponds to the normalization of the two-point function 
of the complete $R$-symmetry current of 
the ${\cal N}=4$ SYM$_4$ \cite{OP,MPet}.

The function $F_2(Y)$ receives contributions 
both from the $R$-symmetry current
and from traceless symmetric rank $2k+1$ 
tensors with approximate dimension $2k+5$.
Since $R_{\mu}^{{\cal J}_{15}}$ is non-renormalized, 
the presence of the function 
$G_2$ shows that operators from the above tensor 
tower acquire anomalous dimensions.
We can find the anomalous dimension of 
the lowest current $O_{15}$ in this tower
whose free field theory counterpart $O_{\mu}^{{\cal J}_{15}}$ with
conformal dimension $\D^{(0)}=5$ was  
discussed in section 3. In fact in perturbation theory
the free-field operator $O_{\mu}^{{\cal J}_{15}}$
contains in the split a descendent of $O_1$  
and currents from the Konishi and the $\Xi$-multiplets.
It is a descendent of $O_{\bf 1}$ that is responsible 
for the logarithmic term in (\ref{4pt15}) and, therefore,
its anomalous dimension at strong coupling is 
$-\frac{16}{N^2}$. 
Comparing 
the coefficient in front of $v^2Y\log v$ in (\ref{4pt15}) with 
the asymptotic (\ref{OPEa}) one gets
$$
\frac{1}{4}\frac{C_{OOO_{15}}^2}{C_{O_{15}}}\D^{(1)}=-\frac{16}{5N^2}
$$ 
and substituting $\D^{(1)}=-\frac{16}{N^2}$ one obtains 
$\frac{C_{OOO_{15}}^2}{C_{O_{15}}}=\frac{4}{5}$ that is different 
from the free-field ratio $\frac{C_{OOO_{15}}^2}{C_{O_{15}}}=4$.
%%%%%%%%%%%%%%%%%%%%%%%%%%%%%%%%%%%%%%%%%%%%%%%%%%%%%%%%%%%%%%%%%%%%%%%%%
\subsection{Projection on ${\bf 175}$}
%%%%%%%%%%%%%%%%%%%%%%%%%%%%%%%%%%%%%%%%%%%%%%%%%%%%%%%%%%%%%%%%%%%%%%%%%
Only double-trace operators transforming in the {\bf 175} appear in the
free field theory OPE (\ref{freeope}) and in the strong coupling
OPE (\ref{strongope}).
Applying the projector $(P_{\bf 175})_{I_1I_2I_3I_4}$ 
constructed in Appendix D 
to the 4-point function we find the following 
expression for the contribution 
of the operators in the ${\bf 175}$:
\bea
\nonumber
&&\left. \langle O^{I_1}(x_1)O^{I_2}(x_2)O^{I_3}(x_3)
O^{I_4}(x_4)\rangle
\right|_{{\bf 175}}=
\frac{8}{\pi^2 N^2}\frac{C^{I_1I_2}_{\J_{175}}C^{I_3I_4}_{\J_{175}}}
{x_{12}^4x_{34}^4}
\biggl[    
-\frac{v^2}{2}\bar{D}_{1212}+\frac{v^2}{2}\bar{D}_{2112} \\
&&+\left(v^2+v^3-\frac{v^3}{u}\right)\bar{D}_{2323} 
+\left(v^2-v^3-\frac{v^3}{u}\right)\bar{D}_{3223} 
\biggr].
\nonumber
\eea
Expanding $\bar{D}$ functions in $v$ we keep 
the leading terms $v^2$ and $v^3$ 
\bea
\nonumber
\left. \langle O^{I_1}(x_1)O^{I_2}(x_2)O^{I_3}
(x_3)O^{I_4}(x_4)\rangle
\right|_{{\bf 175}}&=&
\frac{1}{N^2}\frac{C^{I_1I_2}_{\J_{175}}
C^{I_3I_4}_{\J_{175}}}{x_{12}^4x_{34}^4}
\biggl[v^2F_2(Y)+v^3F_3(Y)\nonumber \\
&&\hspace{1cm}+v^3\log v G_3(Y)\biggr]
\nonumber
\eea
with 
\ba
F_2(Y)&=&-\frac{4(Y(Y-2)+2(Y-1)\log(1-Y))}{Y^2(Y-1)}, \nonumber \\
F_3(Y)&=&\frac{4(28-28Y+3Y^2)}{Y^3(Y-1)}
-\frac{8(38-25Y+Y^2)\log(1-Y)}{Y^4}, \nonumber \\
&+&\frac{96(Y-2)}{Y^4}\Li(Y), \nonumber \\
\nonumber
G_3(Y)&=&\frac{8}{Y^3}\left(\frac{12-12Y+Y^2}{Y-1}
+\frac{6(Y-2)\log(1-Y)}{Y}\right).
\ea
The function $F_2$ receives contributions from 
tensor operators of rank $2k+1$
with approximate dimensions $2k+5$. Since 
the term proportional to $v^2\log v$
is absent, we conclude that these tensor 
operators have protected conformal dimensions. The lowest current
$O_{\mu}^{{\cal J}_{175}}$  among them,
with dimension 5, was  discussed in section 3. 
Note that these operators also contribute to $F_3$ together 
with operators of 
rank $2k+1$ and approximate dimensions $2k+7$.
For the two terms of the $Y$-expansion one finds 
\bea
\label{4pt175}
\left. \langle O^{I_1}(x_1)O^{I_2}(x_2)O^{I_3}(x_3)O^{I_4}(x_4)\rangle
\right|_{{\bf 175}}=
\frac{1}{N^2}\frac{C^{I_1I_2}_{\J_{175}}C^{I_3I_4}_{\J_{175}}}
{x_{12}^4x_{34}^4}
\biggl[-\frac{4}{3}v^2Y-2v^2Y^2\biggr].
\eea
This allows us to determine the $1/N^2$ correction to 
the 2- and 3-point normalization 
constant $C_{O_{175}}$ of the operator $O_{\mu}^{{\cal J}_{175}}$. 
Taking into account that in free field theory 
$C_{OOO_{175}}^{fr}=C_{O_{175}}^{fr}=4$ as can be easily seen
from the free field theory 4-point function (\ref{free4uv}), 
we write as
$$
C_{O_{175}}=4\left(1+\frac{1}{N^2}C_{O_{175}}^{(1)}\right).
$$
Then from the first term of order $v^2$ in (\ref{4pt175}) one finds 
$$
C_{O_{175}}^{(1)}=-\frac{2}{3}.
$$
%Thus, we conclude that the dimension of 
%the operator $O_{\mu}^{{\cal J}_{175}}$
%does not receive anomalous corrections at 
%order $1/N^2$ while for its 2- and 3-point 
%functions this  is not the case.
Apparently, the splitting mechanism is again at work, i.e. the 
corresponding free field theory operator is split in two 
orthogonal parts carrying different representation of the supersymmetry; 
one has protected both its dimension and the 
normalization constants, while the other one is dual to a string
mode and decouples at strong coupling. 

%%%%%%%%%%%%%%%%%%%%%%%%%%%%%%%%%%%%%%%%%%%%%%%%%%%%%%%%%%%%%%%%%%%%%%%%%%%
\section{Conclusions}
%%%%%%%%%%%%%%%%%%%%%%%%%%%%%%%%%%%%%%%%%%%%%%%%%%%%%%%%%%%%%%%%%%%%%%%%%%%
We studied in detail the 4-point functions of the lowest weight
CPOs and we showed that they have a structure 
compatible with the OPE of CPOs predicted by the AdS/CFT
correspondence. We demonstrated that all power-singular terms
in the 4-point functions exactly match the corresponding 
terms in the conformal partial wave amplitudes of
the CPOs, of the $R$-symmetry current and of the stress tensor. 
As these operators are dual to type IIB supergravity fields, we 
concluded that the  
operators dual to string modes, which appear in
the free field theory OPE, decouple in the strong coupling limit.

We also computed the anomalous dimensions and the leading $1/N^2$ 
corrections to the normalization constants of the 2- and 3-point
functions of the scalar double-trace operators with approximate
dimension 4 and 
of vector operators with approximate dimension 5. 
The only scalar double-trace operator that acquires an 
anomalous dimension appears to be the operator in the singlet of 
the $R$-symmetry group $SO(6)$. 
The double-trace operator in the {\bf 20} seems to be 
protected, however as this does not follow from the
shortening condition 
discussed in \cite{AFer,AFer2} we do not have a satisfactory
explanation for such a non-renormalization property.  

The anomalous dimension of the singlet
operator is negative, hence this operator is
relevant and can be used to study non-conformal 
deformations of the $\N =4$ SYM$_4$. 
All other scalar double-trace operators have protected
dimension 4 and are marginal. They can be added to the  
Lagrangian in order to study conformal deformations. 
Nevertheless, it is unclear at present
how dual deformations of type IIB supergravity 
(or string theory) can be described.    

We have also found several towers of traceless symmetric 
double-trace operators in the {\bf 105}, 
{\bf 84} and {\bf 175} irreps, whose anomalous conformal dimensions 
vanish. The rank-$2k$ tensor operators of dimension $6+2k$ 
satisfy the shortening 
condition A') of \cite{AFer2}. However, even if they contain  
the highest weight states of the $SU(2,2|4)$ superalgebra
the shortening condition A') does not imply  
non-renormalization of the corresponding multiplets. On the other hand 
operators from other towers are certainly not the highest weight states,
and at present we are not aware if the lowest weight states of 
their supermultiplets satisfy the shortening condition 
responsible for non-renormalization. 

There are two interesting facts related to the structure of
the leading log-dependent terms in the 4-point functions. 
Namely, all the functions $G(Y)$ which appear in (\ref{4p1}), 
(\ref{4p200}) and so on, differ from each other by
some simple rational factors. We expect that this is an indication
that the anomalous dimensions of all double-trace operators 
may be related by some relatively  simple formula. Then, the leading 
$\log v$-dependent terms appear in the 4-point functions exactly
at the same order of $v$ where the dilogarithm $\Li$ appears for
the first time.  

\vskip 1cm
{\bf ACKNOWLEDGMENTS}
We would like to thank  A. Tseytlin 
for valuable comments.
G.A. is grateful to S. Theisen and, especially, to 
S. Kuzenko for discussions of the structure of the Konishi multiplet. 
S.F. is grateful to S. Mathur and, especially, 
to  A. Tseytlin 
for valuable discussions. A.C.P. wishes to thank 
W. R\"uhl for sharing with him his insight on CFT. 
The work of G.A. was
supported by the Alexander von Humboldt Foundation and in part by the
RFBI grant N99-01-00166. The work of S.F. was supported by
the U.S. Department of Energy under grant No. DE-FG02-96ER40967 and
in part by RFBI grant N99-01-00190. The work of A. C. P. was supported
by the Alexander von Humboldt Foundation. G.A. and A.C.P. wish to
acknowledge the warm hospitality and financial support of the
E.S.I. in Vienna where part of the work was done.

\newpage
%%%%%%%%%%%%%%%%%%%%%%%%%%%%%%%%%%%%%%%%%%%%%%%%%%%%%%%%%%%%%%%%%%%%%%%%%%%
\section{Appendix A.  Free field OPE and conformal blocks}
\setcounter{equation}{0}
%%%%%%%%%%%%%%%%%%%%%%%%%%%%%%%%%%%%%%%%%%%%%%%%%%%%%%%%%%%%%%%%%%%%%%%%%%%
A quasi-primary field of the CFT appearing in the OPE together 
with all its derivative 
descendents is known as a conformal block. If two fields 
$O^{\a}$ and $O^{\b}$ transforming in some representation of an 
$R$-symmetry group have the one and the same conformal dimension $\D$
then their OPE has the following structure 
\bea
\label{OPEaa}
O^{\a}(x)O^{\b}(0)&=&\frac{1}{(x^2)^{\frac{1}{2}(2\D-\D_O)}}
C(x,\partial)O^{\a\b}(0)\\
\nonumber
&+&\frac{1}{(x^2)^{\frac{1}{2}(2\D-\D_J+1)}}C_{\mu}
(x,\partial)J_{\mu}^{\a\b}(0)\\
\nonumber
&+&\frac{1}{(x^2)^{\frac{1}{2}(2\D-\D_T+2)}}
C_{\mu\nu}(x,\partial)T_{\mu\nu}^{\a\b}(0)+....
\eea
Here we identify the leading quasi-primary fields 
with conformal dimensions $\D_O$, $\D_J$ and $\D_T$
as a scalar $O^{\a\b}$, a vector 
current $J_{\mu}^{\a\b}$ and a symmetric traceless
second rank tensor $T_{\mu\nu}^{\a\b}$ respectively. 
The OPE coefficient $C(x,\partial)$ 
denotes a power series in derivatives 
generating the conformal block $[O^{\a\b}]$
of the scalar $O^{\a\b}$. Similarly we denote the OPE coefficients for
for the other fields. 

The structure of the conformal blocks 
is uniquely fixed by the conformal symmetry 
and it may be found by requiring 
consistency of the OPE with 2- and 3-point 
functions of the fields involved.  
In particularly, the conformal block of a 
scalar field with dimension 
$\D$ is given by the following differential operator \cite{FGGP,P}:
\bea
C(x,\partial_y)&=&\frac{1}{B\left(\frac{1}{2}\D_O, \frac{1}{2}\D_O\right)}
\sum_{m=0}^{\infty}\frac{1}{m!(\D_O-\eta+1)_m} \\
\nonumber
&\times &\int_0^1\rmd t[t(1-t)]^{\frac{1}{2}\D_O-1}
\left( -\frac{1}{4}t(1-t)x^2\D_y  \right)^m e^{tx\partial_y},
\eea
where the Euclidean space-time dimension  $d$ 
enters as $d=2\eta$, 
$x\partial_y = x^{\mu}\partial_{y,\mu}$, $\D_y=\partial_y^2$ and we
use the Pochhammer symbol $(a)_n= \Gamma(a+n)/\Gamma(a)$. 
In what follows we need to specify explicitly the first three 
terms of  $C(x,\partial_y)$ in the derivative expansion: 
\bea
C(x,\partial_y)=1+\frac{1}{2}(x\partial_y)
+\frac{\D_O+2}{8(\D_O+1)}(x\partial_y)^2
-\frac{\D_O}{16(\D_O+1)(\D_O+1-\eta )}x^2\D_y+...
\eea
The conformal blocks of a conserved vector current 
and a conserved second rank tensor 
with canonical dimensions $2\eta-1$ and $\eta$ 
respectively are also available.
For a vector current one has \cite{LR}
\bea
C_{\mu}(x,\partial_y)&=&\frac{x_{\mu} }{B(\eta,\eta)}
\sum_{m=0}^{\infty}\frac{1}{m!(\eta)_m}
\int_0^1\rmd t[t(1-t)]^{\eta-1}\left(-\frac{1}{4}t(1-t)x^2\D_y\right)^m e^{tx\partial_y}\\
\nonumber
&=&x_{\mu}+\frac{1}{2}x_{\mu}(x\partial_y) 
+\frac{x_{\mu}}{4(2\eta+1)}
\left((\eta+1)(x\partial_y)^2-\frac{1}{2}x^2\D_y\right)+\ldots 
\eea
and for a conserved symmetric traceless tensor one finds (see
Appendix E) 
\bea
\nonumber
C_{\mu\nu}(x,\partial_y)&=&
\frac{x_{\mu}x_{\nu}}{B(\eta+1,\eta+1)}
\sum_{m=0}^{\infty}\frac{1}{m!(\eta+1)_m}
\int_0^1\rmd t[t(1-t)]^{\eta}\left(-\frac{1}{4}t(1-t)x^2\D_y\right)^m e^{tx\partial_y}\\
\nonumber
&=&x_{\mu}x_{\nu}+\ldots  
\eea

Using the above formulae, one can now consider the operator product
$:O^{\a}(x)O^{\b}(0):$  
in a free field theory
and find explicit expressions for 
$J_{\mu}^{\a\b}$ and $T_{\mu\nu}^{\a\b}$. 
Indeed, from the Taylor expansion 
one sees that the leading component is a quasi-primary field
$O^{\a\b}=:O^{\a}(x)O^{\b}(0):$ 
with conformal dimension $\D_O=2\D$, therefore 
it should appear in the 
OPE with its whole conformal block. Subtracting 
from the Taylor expansion
the first three terms of the conformal block of 
the scalar with dimension $2\D$ 
we find at the next level another quasi-primary 
operator $O_{\mu}^{\a\b}$
that turns out to be a vector current 
$J_{\mu}^{\a\b}=\frac{1}{2}:(\partial_{\mu}O^{\a}O^{\beta}-O^{\a}
\partial_{\mu}O^{\b}):$ with dimension $\D_J=2\D+1$. 
Now subtracting from what we get the first two terms of 
the conformal block of the 
vector current\footnote{We do not assume here that $J_{\mu}^{\a\b}$ is conserved, however the first two terms in the 
conformal blocks of the conserved and 
non-conserved vector currents are the same.} 
and decomposing the resulting second rank tensor on 
the traceless and trace parts we are left with two new fields,
one is a tensor and another one is a new scalar, which are given by 
\bea
T_{\mu\nu}^{\alpha\beta}&=&\frac 12 :(\pa_\mu O^{\alpha}\pa_\nu O^{\beta}+
\pa_\nu O^{\alpha}\pa_\mu O^{\beta}):
-\frac{\D}{2(2\D +1)}\pa_\mu\pa_\nu (:O^{\alpha}O^{\beta}:)\nonumber \\
&+&\frac{\d_{\mu\nu}}{4\eta}\left( 
-\frac{\D +1}{2\D +1}\pa^2 (:O^{\alpha}O^{\beta}:)+
:\pa^2O^{\alpha}O^{\beta}:+:O^{\alpha}\pa^2O^{\beta}:\right), \nonumber \\ 
T^{\alpha\beta}&=&\frac{1}{4\eta }\left( 
-\frac{\D -\eta+1}{2\D+1-\eta}\pa^2 (:O^{\alpha}O^{\beta}:)+
:\pa^2O^{\alpha}O^{\beta}:+:O^{\alpha}\pa^2O^{\beta}:\right).
\nonumber
\eea
The transformation properties of these fields under 
the conformal group show that 
the are both quasi-primary. Thus, for $\eta=2$ 
we get the desired result (\ref{taylor}). Note that 
$T_{\mu\nu}^{\alpha\beta}$ is conserved 
while $T^{\alpha\beta}$ vanishes 
on-shell as soon as $\eta=\D+1$. Clearly 
with the knowledge of the conformal 
blocks of the higher rank tensor operators the procedure 
of identifying the quasi-primary operators on 
the r.h.s of (\ref{biloc}) may be extended
to any desired order. 

\section{Appendix B. Conformal partial wave amplitudes of a scalar, 
a conserved vector current and the stress tensor}
\setcounter{equation}{0}

The full contribution of the conformal block of an operator carrying
and irreducible representation of the conformal group into  the
4-point function  
is known as the conformal partial wave amplitude (CPWA). 
The scalar CPWA was computed in \cite{FGGP} by evaluating 
the corresponding scalar exchange diagram. If we consider operators 
with the same conformal dimension, then the CPWA of a scalar operator with 
dimension $\D_S$ contributes to its 4-point function as \cite{P}:
\bea
\label{CPWAs0}
{\cal H}_S(v,Y)=v^{\frac{\D_S}{2}}\sum_{n=0}^{\infty}\frac{v^n}{n!}
\frac{\left(\frac{1}{2}\D_S\right)_n^4}{(\D_S)_{2n}(\D_S+1-\eta)_n}
~_2F_1\left(\frac{1}{2}\D_S+n,\frac{1}{2}\D_S+n;\D_S+2n;Y\right)\!,
\eea   
where we have represented the result as the convergent series 
in conformal variables $v$ and $Y$. The first few terms of the $v, Y$
expansion of ${\cal H}_S(v,Y)$ are given in (\ref{OPEa}). In particular,
for $\D_S=2$ the first term of $v$-expansion reads as 
\ba
\label{CPWAs}
{\cal H}_S(v,Y)=\frac{3}{40}vF_1(Y)+\ldots \, ,
\ea
where $F_1(Y)$ is defined in section 4.2.

The CPWA of traceless symmetric tensors of dimension $\Delta$ and
rank $l$, corresponding to irreducible representations of dimension
$\D$ and spin $l$ of $SO(d,2)$,  can be also
calculated in CFT as the relevant graphs 
reduce to sums of scalar exchanges. Using the following normalization
prescriptions \cite{Mack,Fradkin}
for the 2- and 3-point functions of the
exchanged tensor fields
\ba
%\label{t2ptf}
\nonumber
\langle M_{\m_1,..,\m_l}(x_1)M_{\n_1,..,\n_l}(x_2)\rangle &=&
C_{\D,l}\frac{{\cal N}(\D,l)}{x_{12}^{2\D}}
\left[\Bigl\{I_{\m_1\n_1}(x_{12}) \cdots
    I_{\m_l\n_l}(x_{12})\Bigl\}_{sym}-{\rm traces} \right]\, , \\
\nonumber
\langle O(x_1)O(x_3)M_{\mu_1,\mu_2,..,\mu_l}(x_5)\rangle 
&=& \frac{g_{\tilde{\Delta}\tilde{\Delta}
    \Delta,l}\,\,{\cal N}(\tD;\D,l)
  }{(x^2_{13})^{\tilde{\Delta}-\frac{1}{2}\Delta}(x_{15}^2  
  x_{35}^2)^{\frac{1}{2}\Delta}} \left[
  \frac{\xi_{\mu_1}\xi_{\mu_2}\cdots
    \xi_{\mu_l}}{(\xi^2)^{\frac{1}{2}l}} -{\rm trace
    \,\,terms}\right]\,,
\ea
where the normalization constants are taken to be 
\ba
\nonumber
&&\hspace{-1cm}{\cal N}(\D,l) =
\frac{2^{\D}\Gamma(\D+l)\Gamma(d-\D-1)}{(2\pi)^{\frac12 
    d}\Gamma(\frac12 d-\D)\Gamma(d-\D+l-1)}, \\
\nonumber
%\label{mvertex}
&&\hspace{-1cm} {\cal N}(\tD;\D,l) =\frac{2^{\tD+\frac12\D+\frac12
    l}}{(2\pi)^{\frac12 d}} \!\left(\frac{\Gamma(\tD+\frac12\D+\frac12
    l-\frac12 d) \Gamma(\tD-\frac12\D+\frac12
    l)\Gamma^2(\frac12\D+\frac12 l)}{ \Gamma(d-\tD-\frac12\D+\frac12
    l)\Gamma(\frac12 d-\tD+\frac12\D+\frac12 l)\Gamma^2(\frac12
    d-\frac12\D+\frac12 l)}\right)^{\frac12}\,
\ea
and
\ba
\nonumber
%\label{ksi1}
&&\xi_{\mu}(1,2;3) \,\,=\,\, \frac{(x_{13})_{\mu}}{x_{13}^2}
-\frac{(x_{23})_{\mu}}{x_{23}^2} \,\,\,\,,\,\,\,\,
\xi^2(1,2;3) \,\,= \,\,\frac{x_{12}^2}{x_{13}^2 x_{23}^2}\,, 
\ea
the  contribution of the tensor field to the 4-point function
of a scalar operator with dimension $\tD$ takes the form
\bea
\label{beta1}
\beta_{\tD}(x_1,x_3;x_2,x_4;\D,l) & =
&\beta_{\tD;\D,l}\frac{1}{(x_{13}^2)^{\tD-\frac12\D}
  (x_{24}^2)^{\tD-\frac12 
    d+\frac12\D}} \nonumber \\
& & \times{}\int\rmd^d
x_{\5}\frac{\Bigl\{e_{\m_1}\cdots e_{\m_l}-{\rm
    traces}\Bigl\}\Bigl\{e'_{\m_1}\cdots e'_{\m_l}-{\rm
    traces}\Bigl\}}{(x_{1\5}^2
  x_{3\5}^2)^{\frac12\D}(x_{2\5}^2x_{4\5}^2)^{\frac12 
    d-\frac12\D}}\, .
\eea
The constant $\beta_{\tD;\D,l}$ is then given by 
\bea
\nonumber
\beta_{\tD;\D,l} &= & \frac{g^2_{\tD\tD\D,l}}{C_{\D,l}}
\frac{2^{2\tD+\frac12 d+\frac12 
    l}\Gamma(\tD-\frac12\D +\frac12 l)\Gamma(\tD+\frac12\D+\frac12
  l-\frac12 d)}{(2\pi)^{\frac{d}{2}} \Gamma(\frac12 d-\tD+\frac12\D+\frac12
  l)\Gamma(d-\tD-\frac12\D+\frac12 l)}\,,
\ea
where we have introduced the concise notation
\ba
\nonumber
e_{\m}&=&\frac{\xi_{\m}(1,3;\5)}{|\xi^2(1,3;\5)|^{\frac12}}\, 
\,,\,\,\, e'_{\m}\,\,=\,\,
\frac{\xi_{\m}(2,4;\5)}{|\xi^2(2,4;\5)|^{\frac12}}\,\,,\,\,\,e\cdot  
e\,\,=\,\,e'\cdot  
e'\,\,=\,\,1\,.
\ea
One can show that for the general tensor exchange
(\ref{beta1}) is reduced to a finite sum of
four-star integrals $S(a_1,a_2;a_3,a_4)$: 
\beq
\label{4*}
S(a_1,a_2;a_3,a_4)= \int\rmd^4x_{\5}\frac{1}{x_{1\5}^{2a_1}x_{2\5}^{2a_2}
  x_{3\5}^{2a_3} x_{4\5}^{2a_4}} \, ,
\eeq
which can be directly evaluated. The final result
is obtained after dropping the ``shadow series'' of the four-star
integral, as the latter corresponds to the exchange of the ``shadow tensor''
field with dimension $d-\D$. 

Here, we apply the general formula (\ref{beta1}) to the two cases we are 
interested in the paper; the case of the conserved vector
current with $\D=d-1$ and $l=1$ and the stress tensor with
$\D=d$ and $l=2$.
Choosing to work directly in $d=4$, the contribution of a conserved
vector field in the scalar four-point function is given by
\ba
\label{vcontr}
\b_{2}(x_1,x_2;x_3,x_4;3,1) &=& \b_{2;3,1}\frac{1}{(x_{12}^2)^{2-\frac32}
    (x_{34}^2)^{2-\frac12}}\int\rmd^4 x_{\5}\frac{e\cdot e'}{(x_{1\5}^2
    x_{2\5}^2)^{\frac32}(x_{3\5}^2x_{4\5}^2)^{\frac12}} \, . 
\ea
The inner product $e\cdot e'$ can be written as
\beq
\label{ee'}
\nonumber
e\cdot e' = \frac12\left(\frac{x_{1\5}^2 x_{2\5}^2 x_{3\5}^2
      x_{4\5}^2}{x_{12}^2 x_{34}^2}\right)^{\frac12}\left[
  \frac{x_{24}^2}{ x_{2\5}^2 x_{4\5}^2} -\frac{x_{14}^2}{x_{1\5}^2
    x_{4\5}^2} +\frac{x_{13}^2}{x_{1\5}^2x_{3\5}^2}
  -\frac{x_{23}^2}{x_{2\5}^2 x_{3\5}^2}\right] \,. 
\eeq
Substituting (\ref{ee'}) into (\ref{vcontr}) we obtain four 4-star
functions as
\ba
\label{v4stars}
&&\b_{2}(x_1,x_2;x_3,x_4;3,1) = \frac12
\b_{2;3,1}\frac{1}{(x_{12}^2)^{2-1} 
    (x_{34}^2)^{2}}\times \\
\nonumber
&&\hspace{2cm} \Bigl[ x_{24}^2 S\left(1+\e,2+\e;-\e,1-\e\right)
-x_{14}^2 S\left(2+\e,1+\e;-\e,1-\e\right) \\
\nonumber
&&\hspace{2cm} +x_{13}^2  S\left(2+\e,1+\e;1-\e,-\e\right)
-x_{23}^2 S\left(1+\e,2+\e;1-\e,-\e\right)\Bigl]\, .
\ea
Note that we have also regularized the dimension of the vector field as
$\D=3+2\e$ to deal with the singularities contained in the four-star
functions involved into 
(\ref{v4stars}). The singularities are avoided by keeping the 
regulating parameter $\e$ non-zero in the intermediate stages of the
calculation. The analyticity of the exchange graph
then ensures that taking the limit $\e\rightarrow 0$ at the end of the
calculation one 
recovers the correct result.  Using the expression for the four-star
function derived in \cite{HPR2} we then obtain (here we
present the formula for general $d$ and $\D$ to ensure a wider
applicability of our result)
\ba
\b_{\tilde{\Delta}}(x_1,x_2;x_3,x_4;\Delta,1) &=&-\frac{2}{3}
\b_{\tilde{\Delta};\Delta,1}
\frac{\pi^{\eta}\Gamma(\frac{\D}{2}-\frac12) 
  \Gamma(\frac{\D}{2} +\frac12) 
  \Gamma(\eta-\Delta) }{\Gamma(\eta-\frac{\D}{2}-\frac12 )
  \Gamma(\eta-\frac{\D}{2}+\frac12)\Gamma(\D)}
\frac{1}{(x_{12}^2x_{34}^2)^{\tilde{\D}}}
\nonumber \\
\nonumber
&\times& \left({\cal H}_V(v,Y)+\mbox{shadow part}\right)\, ,\ea  
where the function ${\cal H}_V(v,Y)$ represents the CPWA of the vector current 
\ba
{\cal H}_V(v,Y)&=&-\frac34 v^{\frac{\D-1}{2}}
\sum_{n,m=0}^{\infty} \frac{v^n Y^m}{n!m!}
\frac{1}{(1-\eta+\D)_{n} (\D)_{2n+m}} \\ 
&\times&
\Biggl[\left(\frac{\D
    +1}{2}\right)_{n}^2 \left(\frac{\D-1}{2}\right)_{n+m}^2 \nonumber
+ \left(\frac{\D
    -1}{2}\right)_{n}^2 \left(\frac{\D+1}{2}\right)_{n+m}^2\nonumber
\\
&& \hspace{1cm}- 2\left(\frac{\D
    -1}{2}\right)_{n} \left(\frac{\D
    +1}{2}\right)_{n}\left(\frac{\D+1}{2}\right)_{n+m}
\left(\frac{\D-1}{2}\right)_{n+m} \nonumber \\
\nonumber
&&\hspace{1cm} -Y \left(
\frac{\D-1}{2}\right)_{n}^2 \left(\frac{\D+1}{2}\right)_{n+m}^2\Biggl]\, .
\ea  
For $\D=d-1=3$ the CPWA of the vector current simplifies to give
\ba
\label{cavc0}
{\cal H}_V(v,Y)&=&-\frac34 v
\sum_{n,m=0}^{\infty} \frac{v^n Y^m}{n!m!}
\frac{1}{(2)_{n}(3)_{2n+m}}  \\ 
\nonumber
&\times&
\Biggl[(2)_{n}^2 (1)_{n+m}^2+ (1-Y)(1)_{n}^2(2)_{n+m}^2
- 2(1)_{n}(2)_{n}(2)_{n+m}(1)_{n+m} \Biggl]\,,
\ea  
and it is normalized to start as ${\cal H}_V(v,Y)=\frac{1}{2}Y+\ldots$ (cf. (\ref{OPEa})). 
To make a comparison with the supergravity results in section 4.5 we
need to single  
out in eq.(\ref{cavc0}) the leading-$v$ contribution. Putting in the
previous formula  
$n=0$ and performing the summation in $m$ we obtain
\ba
\label{cavc}
{\cal H}_V(v,Y)&=&\frac{3}{16}vF_1(Y)+\ldots \,, 
\ea   
where $F_1(Y)$ is defined in section 4.5.

Analogously, the contribution of the stress tensor is given by
\ba
\label{tcontr}
\b_{2}(x_1,x_2;x_3,x_4;4,2) &=& \b_{2;4,2}\frac{1}{
    (x_{34}^2)^{2}}\int\rmd^4
  x_{\5}\frac{(e_{\m}e_{\n}-\frac14\d_{\m\n})
    (e_{\m}'e_{\n}'-\frac14\d_{\m\n})}{(x_{1\5}^2 
    x_{2\5}^2)^{2+\e}(x_{3\5}^2x_{4\5}^2)^{-\e}}\, . 
\ea 
Using then (\ref{ee'}) and regularizing the tensor dimension as
$\D=4+2\e$ we obtain
\ba
\label{t4stars}
\b_{2}(x_1,x_2;x_3,x_4;4,2) & =  & \frac{1}{12}\b_{2;4,2}\frac{1}{
    (x_{34}^2)^{2}}\Biggl[\frac{1}{x_{12}^2 x_{34}^2} \Biggl( x_{24}^4
  S(1+\e,3+\e;-1-\e,1-\e) \nonumber \\
&& \hspace{-2cm}+x_{14}^4 S(3+\e,1+\e;-1-\e,1-\e) + x_{13}^4
S(3+\e,1+\e;1-\e,-1-\e) \nonumber \\
 & & \hspace{-2cm}+x_{23}^4 S(1+\e,3+\e;1-\e,-1-\e) -2x_{24}^2 x_{14}^2
 S(2+\e,2+\e;-1-\e,1-\e) \nonumber \\
&& \hspace{-2cm}-2x_{13}^2 x_{23}^2
 S(2+\e,2+\e;1-\e,-1-\e) +2x_{24}^2x_{13}^2  S(2+\e,2+\e;-\e,-\e)
 \nonumber \\
&& \hspace{-2cm}-2x_{24}^2 x_{23}^2
S(1+\e,3+\e;-\e,-\e) -2x_{14}^2 x_{13}^2 S(3+\e,1+\e;-\e,-\e)
\nonumber \\ &&\hspace{-2cm}+2x_{14}^2 x_{23}^2
S(2+\e,2+\e;-\e,-\e)\Biggl) - S(2+\e,2+\e;-\e,-\e)\Biggl]\, .
\ea
One then observes that (\ref{t4stars}) contains
a number of four-star functions which are $O(\e)$ and therefore 
vanish in the $\e\rightarrow 0$ limit. These are all the four-star
functions with $-\e$ in the last two positions. Then, by virtue of  
\beq
\frac{\Gamma(-2-2\e)}{\Gamma(-1-\e)}=-\frac14 +O(\e)\,,
\eeq
the remaining four-star functions give a finite result which reads
\ba
\nonumber
\b_{2}(x_1,x_2;x_3,x_4;4,2) &\sim &
%\frac{\pi^2}{12}\b_{2,4,2}
\frac{1}{x_{12}^4x_{34}^4}
\left({\cal H}_T(v,Y)+\mbox{shadow part}\right), 
\ea
where ${\cal H}_T(v,Y)$ represents the CPWA of the stress tensor:
\ba
\label{cast0}
{\cal H}_T(v,Y)&=&\frac{5}{4}
v\sum_{n,m=0}^{ \infty} \frac{v^nY^m}{n!m!}\frac{1}{(3)_n (4)_{2n+m}}\\
\nonumber
&\times&\Bigl[ (3)^2_n(1)^2_{n+m}+2(3)_n(1)_n(3)_{n+m}(1)_{n+m}
+(1-Y)^2(1)^2_n (3)^2_{n+m} \\
\nonumber
&&-2(3)_n(2)_n(1)_{n+m}(2)_{n+m} -2(1-Y)
(1)_n(2)_n(3)_{n+m} (2)_{n+m}\Bigl]\, .
\ea
The normalization of ${\cal H}_T(v,Y)$ is fixed such that its $v,Y$ expansion
reproduces the corresponding terms in (\ref{OPEa}). 
Again to establish a link with supergravity results in section 4.1 we single 
out the $v$ term in eq.(\ref{cast0}) and, performing the summation in $m$,  get
\ba
\label{cast}
{\cal H}_T(v,Y)=\frac{45}{8}vF_1(Y)+\ldots \, , 
\ea
where $F_1(Y)$ is the function defined in section 4.1. 
This completes the construction of the CPWA for conserved vector and tensor currents. 
%%%%%%%%%%%%%%%%%%%%%%%%%%%%%%%%%%%%%%%%%%%%%%%%%%%%%%%%%%%%%%%%%%%
\section{Appendix C. Series representation for $\bar{D}$-functions}
\setcounter{equation}{0}
%%%%%%%%%%%%%%%%%%%%%%%%%%%%%%%%%%%%%%%%%%%%%%%%%%%%%%%%%%%%%%%%%%%
Here we derive a representation for the 
$\bar{D}_{\D_1\D_2\D_3\D_4}$-functions
in a form of a convergent series in $v$ and 
$Y$ variables by using a technique 
similar to \cite{L}.  

We start with the definition (\ref{D1}). 
Standard Feynman parameter manipulations 
based on the formula 
$$
\frac{1}{z^{\lambda}}=\frac{1}{\Gamma(\lambda)}\int_0^{\infty}\rmd tt^{\lambda-1}e^{-tz},
$$ 
and two integrals 
\bea
\nonumber
&&\int e^{-\sum_it_ix_0^2}x_0^{-d-1+\sum 
\Delta_i}\rmd x_0=\frac{1}{2}
(S_t)^{\frac{d-\sum_i \Delta_i}{2}}\Gamma
\left(\frac{\sum \Delta_i}{2} -
\frac{d}{2} \right), \\
\nonumber
&&\int \rmd^d\vec{x}e^{-t_i|\vec{x}-\vec{x}_i|^2}=
\frac{\pi^{d/2}}{S_t^{d/2}}
e^{-\frac{1}{S_t}\sum_{i< j}t_it_jx_{ij}^2},
\eea
lead to
\ba
&&D_{\D_1\D_2\D_3\D_4}(x_1,x_2,x_3,x_4) \nonumber \\
\nonumber
&&\hspace{0.5cm}=K\int_0^{\infty}\rmd t_1..\rmd
t_4\,t_1^{\D_1-1}\cdots t_4^{\D_4-1} 
\left(S_t\right)^{-\frac{\D_1+\D_2+\D_3+\D_4}{2}} {\rm
  exp}\Biggl[-\frac{1}{S_t}(t_1t_2x_{12}^2+..+t_3 t_4x_{34}^2)\Biggl],
\eea
where the short-hand notations 
\bea
\nonumber
S_t \,=\,t_1+t_2+t_3+t_4 \,,\,\,\,\,\,\,\,K
\,=\,\frac{\pi^{\frac{d}{2}}\Gamma(\frac{
    \D_1+\D_2+\D_3+\D_4}{2}-\frac{d}{2})}{
  2\Gamma(\D_1)\cdots\Gamma(\D_4)}, 
\ea
were introduced. Performing the change of variables
$$
t_i=S_t^{1/2}t_i'=(\sum_i t_j')t_i'\equiv u t_i'\, , \,\,\,\,\,
\det\left(\frac{\partial t_i}{\partial t_j'}\right)=2u^4,
$$
one obtains
\bea
\nonumber
D_{\D_1\D_2\D_3\D_4}(x_1,x_2,x_3,x_4)=
2 K\int_0^{\infty}\rmd t_1..\rmd
t_4\,t_1^{\D_1-1}\cdots t_4^{\D_4-1} 
{\rm  exp}\Biggl[-\sum_{i< j}t_it_jx_{ij}^2\Biggl].
\eea
Now we rescale the variables $t_i$: 
$t_i\to \lambda_i t_i$, where the constant 
parameters $\lambda_i$ are chosen to induce 
the following scale transformations
\bea
\nonumber
t_1t_2\to \frac{1}{x_{12}^2}t_1t_2,\,\,\,
t_1t_3\to \frac{1}{x_{13}^2}t_1t_3,\,\,\,
t_1t_4\to \frac{1}{x_{14}^2}t_1t_4,\,\,\,
t_2t_3\to \frac{1}{x_{23}^2}t_2t_3,
\eea
and as the consequence
\bea
\nonumber
t_2t_4=\frac{t_2t_3\cdot t_1t_4}{t_1t_3}\to 
\frac{x_{13}^2}{x_{14}^2x_{23}^2}
t_2t_4,\,\,\,
t_3t_4=\frac{t_2t_3\cdot t_1t_4}{t_1t_2}\to 
\frac{x_{12}^2}{x_{14}^2x_{23}^2}
t_3t_4.  
\eea
Under this rescaling the integral transforms into 
\ba
&&\hspace{-2cm}D_{\D_1\D_2\D_3\D_4}(x_1,x_2,x_3,x_4)=\nonumber \\
&&=\frac{\bar{D}_{\D_1\D_2\D_3\D_4}(v,Y)  }
{(x_{12}^2)^{\frac{\D_1+\D_2-\D_3-\D_4}{2}}
 (x_{13}^2)^{\frac{\D_1+\D_3-\D_2-\D_4}{2}}
 (x_{23}^2)^{\frac{\D_2+\D_3+\D_4-\D_1}{2}} 
 (x_{14}^2)^{\D_4}},
\ea
where 
\bea
\nonumber
&&\bar{D}_{\D_1\D_2\D_3\D_4}(v,Y)=\\
\nonumber
&&\hskip 0.5cm 2K\int \rmd t_1...\rmd t_4 t_1^{\D_1-1}t_2^{\D_2-1}t_3^{\D_3-1}t_4^{\D_4-1}
{\rm  exp}\Biggl[-t_1t_2-t_1t_3-t_1t_4-t_2t_3-
\frac{v}{u}t_2t_4-v t_3t_4\Biggl],
\eea
and the integral is understood as a function of 
the conformal variables
$v$ and $Y$.

Next, using the Mellin-Barnes integral representation
\bea
\nonumber
{\rm exp}\left[-z\right] = \frac{1}{2\pi{\rm i}}\int_{r-{\rm
      i}\infty}^{r+{\rm i}\infty}\rmd s\,\Gamma(-s)z^s\,,\,\,\,\,\,
    r<0,\,\,\,\,|{\rm arg}\,z| < \sfrac12 \pi,
\eea
for the two exponentials in the last formula which 
involve $\frac{v}{u}$ and
$v$ the integral reduces to 
\bea
\nonumber
&&\bar{D}_{\D_1\D_2\D_3\D_4}(v,Y) =2K 
\int\frac{\rmd \l\,\rmd s}{(2\pi{\rm i})^2} \Gamma(-s) \Gamma(-\l)
v^{\l}\left(\frac{v}{u}\right)^s \nonumber \\
&&\times \int \rmd t_1...\rmd t_4 t_1^{\D_1-1}t_2^{\D_2+s-1}t_3^{\D_3+\l-1}
t_4^{\D_4+s+\l-1}
{\rm  exp}
\Biggl[-t_1t_2-t_1t_3-t_1t_4-t_2t_3\Biggl].
\nonumber
\eea
The following change of variables: 
$$
t_1t_2=u_1,~~~t_1t_3=u_2,~~~t_1t_4=u_3,~~~t_2t_3=u_4,~~~
\det\left( \frac{\partial t_i}{\partial u_j} \right)=\frac{1}{2u_1u_2},
$$
allows one to perform the $t$-integration with the result
\bea
&&\bar{D}_{\D_1\D_2\D_3\D_4}(v,Y) =\nonumber\\
&&=K\int\frac{\rmd \l\,\rmd
s}{(2\pi{\rm i})^2} \,\Biggl[ \Gamma(-s) \Gamma(-\l) 
\Gamma(\sfrac{\D_1+\D_2-\D_3-\D_4}{2}-\l)
\Gamma(\sfrac{ \D_1+\D_3-\D_2-\D_4}{2}-s)\nonumber \\
&&\hspace{4cm}\times
\Gamma(\sfrac{\D_2+\D_3+\D_4-\D_1}{2} +s+\l)
\Gamma(\D_4+s+\l)\,\,v^{\l}\, \left(\frac{v}{u}\right)^s\Biggl].
\nonumber
\eea
The $s$-integration is then performed  by 
using the integral and series representations 
for the hypergeometric function $F(a,b,c;1-z)$:
\bea
\nonumber
F(a,b,c;1-z)&=&\frac{\G(c)}{\G(a)\G(b)\G(c-a)\G(c-b)}\\
\nonumber
&\times&\frac{1}{2\pi i}
\int_{-i\infty}^{i\infty}\rmd s z^{s}\G(-s)\G(c-a-b-s)\G(a+s)\G(b+s),
\eea
and 
\bea
\nonumber
F(a,b,c;1-z)=\frac{\G(c)}{\G(a)\G(b)}\sum_{m=0}^{\infty}
\frac{\G(a+m)\G(b+m)}{\G(c+m)m!}(1-z)^m,
\eea
where one needs to substitute
$$
a=\frac{\D_2+\D_3+\D_4-\D_1}{2}+\l ,~~~b=\D_4+\l ,~~~c=\D_3+\D_4+2\l .
$$
Thus one arrives at the convergent hypergeometric series in the variable $Y$:
\ba
&&\bar{D}_{\D_1\D_2\D_3\D_4}(v,Y) =\nonumber\\
&&=K\sum_{m=0}^{\infty}\frac{Y^m}{m!} \Biggl\{\int
\frac{\rmd\l}{2\pi{\rm i}}\,\Biggl[\Gamma(-\l) \frac{
  \Gamma(\frac{\D_1+\D_2-\D_3-\D_4}{2}-\l)
  \Gamma(\frac{\D_3+\D_4+\D_1-\D_2}{2}+\l)
  \Gamma(\D_3+\l)}{\Gamma(\D_3+\D_4+2\l+m)} \nonumber\\
&&
\hspace{4cm}\Gamma(\sfrac{\D_2+\D_3+\D_4-\D_1}{2}+\l+m)\Gamma(\D_4+\l+m)
\,v^{\l}\Biggl]\Biggl\}.
\ea 
Since for any $\bar{D}$-function occurring in the 4-point function of CPOs
the quantity $\D_1+\D_2-\D_3-\D_4$ is an integer, the final Mellin-Barnes
integral receives a 
contribution from double poles and, therefore, the  
integration can be done by  using the general formula
\beq
\int_{\cal C}\frac{\rmd s}{2\pi{\rm i}} \Gamma^2(-s)\,g(s)\,v^s =
\sum_{n=0}^{\infty} \frac{v^n}{(n!)^2}\left[ 2\psi(n+1)g(n)-g(n)\ln
  v-\frac{\rmd }{\rmd \xi}[g(\xi)]_{\xi=n}\right],
\eeq
valid for any function $g(s)$ regular at $s=0$. In this way we arrive at the 
representation for $\bar{D}$-functions in terms of double convergent series 
in $v$ and $Y$ variables.

Below we list explicitly the series representations 
for $\bar{D}$-functions we used in the paper
\ba
\nonumber
\bar{D}_{2222}(v,Y)&=&\pi^2\sum_{m=0}^{\infty}\frac{Y^m}{m!}\frac{v^n}{(n!)^2}
 \frac{\Gamma(n+2)^2\Gamma(2+n+m)^2}{\Gamma(4+2n+m)}\\
&\times & \left(
-\frac{1}{n+1}+\psi(4+2n+m)-\psi(n+m+2)-\frac{1}{2}\ln{v} 
\right),  \nonumber \\
\nonumber      \\
%%%%%%%%%%%%%%%%%%%%%%%%%%%%%%%%%%%%%%%%%%%%%%%%%%%%%%%%%%%%%%
\nonumber
\bar{D}_{2112}(v,Y)&=&\frac{\pi^2}{2}
\sum_{m=0}^{\infty}\frac{Y^m}{m!}\frac{v^n}{(n!)^2}
\frac{\Gamma(n+2)\Gamma(n+1)\Gamma(n+m+1)\Gamma(n+m+2)}{\Gamma(3+2n+m)}
\\
\nonumber
&\times & \left(
-\frac{1}{n+1}+2\psi(3+2n+m)-\psi(n+m+1)-\psi(n+m+2)-\ln{v} 
\right),  \\
\nonumber      \\
%%%%%%%%%%%%%%%%%%%%%%%%%%%%%%%%%%%%%%%%%%%%%%%%%%%%%%%%%%%%%%
\nonumber
\bar{D}_{1212}(v,Y)&=&\pi^2
\sum_{m=0}^{\infty}\frac{Y^m}{m!}\frac{v^n}{(n!)^2}
 \frac{\Gamma(n+1)^2\Gamma(n+m+2)^2}{\Gamma(3+2n+m)}\\
\nonumber
&\times & \left(\psi(3+2n+m)-\psi(n+m+2)-\frac{1}{2}\ln{v} 
\right),   \\
\nonumber      \\
%%%%%%%%%%%%%%%%%%%%%%%%%%%%%%%%%%%%%%%%%%%%%%%%%%%%%%%%%%%%%%%
\nonumber
\bar{D}_{2211}(v,Y)&=&-\frac{\pi^2}{2}
\sum_{m=0}^{\infty}\frac{Y^m}{m!}\frac{v^n}{(n!)^2}
 \frac{n\Gamma(n+1)^2\Gamma(n+m+1)^2}{\Gamma(2+2n+m)}\\
\nonumber
&\times & \left(-\frac{1}{n}-2\psi(n+m+1)+2\psi(2+2n+m)-\ln{v} 
\right), \\
\nonumber      \\
%%%%%%%%%%%%%%%%%%%%%%%%%%%%%%%%%%%%%%%%%%%%%%%%%%%%%%%%%%%%%%%%
\nonumber  
\bar{D}_{3322}(v,Y)&=&-\frac{\pi^2}{4}
\sum_{m=0}^{\infty}\frac{Y^m}{m!}\frac{v^n}{(n!)^2}
 \frac{n\Gamma(n+2)^2\Gamma(2+n+m)^2}{\Gamma(4+2n+m)}
\\
\nonumber
&\times & \left(-\frac{3n+1}{n(n+1)}+2\psi(4+2n+m)-2\psi(2+n+m)-\ln{v} 
\right),  \\
\nonumber      \\
%%%%%%%%%%%%%%%%%%%%%%%%%%%%%%%%%%%%%%%%%%%%%%%%%%%%%%%%%%%%%%%%%
\nonumber  
\bar{D}_{2323}(v,Y)&=&\frac{\pi^2}{2}
\sum_{m=0}^{\infty}\frac{Y^m}{m!}\frac{v^n}{(n!)^2}
\frac{\Gamma(n+2)^2\Gamma(3+n+m)^2}{\Gamma(5+2n+m)}
\\
\nonumber
&\times & \left(-\frac{1}{n+1}+\psi(5+2n+m)-\psi(3+n+m)-\frac{1}{2}\ln{v} 
\right), \\
\nonumber      \\
%%%%%%%%%%%%%%%%%%%%%%%%%%%%%%%%%%%%%%%%%%%%%%%%%%%%%%%%%%%%%%%%%%%%
\nonumber   
\bar{D}_{3223}(v,Y)&=&\frac{\pi^2}{4}
\sum_{m=0}^{\infty}\frac{Y^m}{m!}\frac{v^n}{(n!)^2}
\frac{\Gamma(n+2)\Gamma(n+3)\Gamma(2+n+m)\Gamma(3+n+m)}{\Gamma(5+2n+m)}\\
\nonumber
&\times&  \left(-\frac{3n+5}{(n+1)(n+2)}+2\psi(5+2n+m)\right. \nonumber \\
&-&\psi(2+n+m)-\psi(3+n+m)-\ln{v} 
\left. \right) .
\eea

\section{Appendix D. Projectors}
\setcounter{equation}{0}
Here we give an explicit construction of the 
projectors that single out the contributions of 
irreps occurring in the decomposition 
${\bf 20}\times {\bf 20}$ of $SO(6)$ from the 
4-point function of the lowest weight CPOs.

Matrices $C_{ij}^I$ and $C_{ij}^{{\cal J}_{15}}$ 
introduced in section 3 obey 
to the following summation formulae \cite{AF7}:
\bea
\nonumber
\sum_{I}C_{ij}^IC_{kl}^I=\frac{1}{2}\d_{ik}\d_{jl}
+\frac{1}{2}\d_{il}\d_{jk}
-\frac{1}{6}\d_{ij}\d_{kl},~~~~
\sum_{{\cal J}_{15}}C_{ij}^{{\cal J}_{15}}C_{kl}^{{\cal J}_{15}}
=\frac{1}{2}(\d_{ik}\d_{jl}-\d_{il}\d_{jk}).
\eea
It is then easy to check that the orthonormal Clebsh-Gordon 
coefficients $C^{I_1I_2}_{{\cal J}_{20}}$
and $C^{I_1I_2}_{{\cal J}_{15}}$ are given by 
\bea
\label{20and15}
C^{I_1I_2}_{{\cal J}_{20}}=\frac{3^{1/2}}{5^{1/2}}C^{I_1I_2I},~~~~
C^{I_1I_2}_{{\cal J}_{15}}=\frac{1}{2^{1/2}}C^{I_1}_{ij}C^{I_2}_{jk}
C_{ik}^{{\cal J}_{15}}.
\eea
The other coefficients are constructed in a similar 
manner. Irreps $\bf{84}$,   $\bf{105}$
and $\bf{175}$ are described by traceless 
rank 4 tensors $C^{{\cal J}_D}_{ijkl}$
with the normalization condition
$$
C^{{\cal J}_D}_{ijkl}C^{{\cal J}'_{D}}_{ijkl}=\d^{{\cal J}_{D}{\cal J}'_{D}}.
$$  
Tensor $C^{{\cal J}_{84}}_{ijkl}$ is 
antisymmetric in $i,k$ and in $j,l$
and symmetric under permutation of the 
pairs $ij$ and $kl$. It is also required to obey 
the condition $\eps_{ijklmn}C^{{\cal J}_{84}}_{klmn}=0$. Then
$C^{{\cal J}_{105}}_{ijkl}$ is a totally symmetric and, finally, 
$C^{{\cal J}_{175}}_{ijkl}$ is 
symmetric in $i,k$ and in $j,l$ and antisymmetric 
under permutation of the pairs $ij$ and $kl$.

A projector on the contribution of irrep ${\bf D}$ 
into the 4-point function is 
defined by (\ref{projD}) with $\nu_D$ being 
the dimension of the representation.
The sums $C^{I_1I_2}_{{\cal J}_{20}}C^{I_1I_2}_{{\cal J}_{20}}$ and 
$C^{I_1I_2}_{{\cal J}_{15}}C^{I_1I_2}_{{\cal J}_{15}}$ 
are computed straightforwardly 
by using eqs.(\ref{20and15}). To find the other 
projectors we introduced the following 
three tensors $Q^{I_1I_2}_{\bf D}$ 
being elements of the corresponding representations:
\bea
\nonumber
(Q^{I_1I_2}_{\bf 84})_{ijkl}&=&
C_{ij}^{I_1}C_{kl}^{I_2}-C_{kj}^{I_1}C_{il}^{I_2}
+C_{kl}^{I_1}C_{ij}^{I_2}-C_{il}^{I_1}C_{kj}^{I_2}\\
\nonumber
&+&\frac{1}{4}(C_{im}^{I_1}C_{mj}^{I_2}\d_{kl}-
C_{km}^{I_1}C_{mj}^{I_2}\d_{il}
+C_{km}^{I_1}C_{ml}^{I_2}\d_{ij}-C_{im}^{I_1}C_{ml}^{I_2}\d_{kj}) 
\\
\nonumber
&+&\frac{1}{4}(C_{jm}^{I_1}C_{mi}^{I_2}\d_{kl}-
C_{jm}^{I_1}C_{mk}^{I_2}\d_{il}
+C_{lm}^{I_1}C_{mk}^{I_2}\d_{ij}-C_{lm}^{I_1}C_{mi}^{I_2}\d_{kj}) 
\\
\nonumber
&-&\frac{1}{10}\d^{I_1I_2}(\d_{ij}\d_{kl}-\d_{il}\d_{kj}),
\eea
%%%%%%%%%%%%%
\bea
\nonumber
(Q_{\bf 105}^{I_1I_2})_{ijkl}&=&C_{ij}^{I_1}C_{kl}^{I_2}+
C_{ik}^{I_1}C_{jl}^{I_2}
+C_{il}^{I_1}C_{jk}^{I_2}
+C_{kl}^{I_1}C_{ij}^{I_2}+C_{kj}^{I_1}C_{il}^{I_2}
+C_{jl}^{I_1}C_{ik}^{I_2}\\
\nonumber
&-&\frac15\d_{ij}(C_{km}^{I_1}C_{ml}^{I_2}
+ C_{lm}^{I_1}C_{mk}^{I_2})
-\frac15\d_{kl}(C_{im}^{I_1}C_{mj}^{I_2}
+ C_{jm}^{I_1}C_{mi}^{I_2})\nonumber \\
&-&\frac15\d_{ik}(C_{jm}^{I_1}C_{ml}^{I_2}
+ C_{lm}^{I_1}C_{mj}^{I_2})-\frac15\d_{il}(C_{jm}^{I_1}C_{mk}^{I_2}\nonumber \\
&+& C_{km}^{I_1}C_{mj}^{I_2})
-\frac15\d_{jk}(C_{im}^{I_1}C_{ml}^{I_2}+ C_{lm}^{I_1}C_{mi}^{I_2})
-\frac15\d_{jl}(C_{im}^{I_1}C_{mk}^{I_2}+ C_{km}^{I_1}C_{mi}^{I_2})\\
\nonumber
&+&\frac{1}{20}(\d_{ij}\d_{kl}+\d_{ik}\d_{jl}+\d_{il}\d_{jk})\d^{I_1I_2},
\eea
%%%%%%%%%%%%%%%
\bea
\nonumber
(Q^{I_1I_2}_{\bf 175})_{ijkl}&=&C_{ik}^{I_1}C_{jl}^{I_2}-C_{jl}^{I_1}C_{ik}^{I_2}
-\frac{1}{8}\d_{ij}(C_{km}^{I_1}C_{ml}^{I_2}-C_{lm}^{I_1}C_{mk}^{I_2})
-\frac{1}{8}\d_{kj}(C_{im}^{I_1}C_{ml}^{I_2}-C_{lm}^{I_1}C_{mi}^{I_2})\\
\nonumber
&-&\frac{1}{8}\d_{il}(C_{km}^{I_1}C_{mj}^{I_2}-C_{jm}^{I_1}C_{mk}^{I_2})
-\frac{1}{8}\d_{kl}(C_{im}^{I_1}C_{mj}^{I_2}-C_{jm}^{I_1}C_{mi}^{I_2}).
\eea
%%%%%%%%%%%%%
Clearly one may write
\bea
\nonumber
C_{{\cal J}_D}^{I_1I_2}C_{ijkl}^{{\cal J}_D}=\gamma_D (Q_{{\bf D}}^{I_1I_2})_{ijkl},
\eea
where $\gamma_D$ is a normalization constant. Then one finds
\bea
\nonumber
C_{{\cal J}_D}^{I_1I_2}C_{{\cal J}_D}^{I_3I_4}=
C_{{\cal J}_D}^{I_1I_2}C_{ijkl}^{{\cal J}_D}C_{{\cal J}'_D}^{I_3I_4}C_{ijkl}^{{\cal J}_D'}=\
\gamma_D^2(Q_{{\bf D}}^{I_1I_2})_{ijkl}(Q_{{\bf D}}^{I_3I_4})_{ijkl}
\eea
with the normalization constant $\gamma_D$ following from
$$
\nu_D=\gamma_D^2(Q_{{\bf D}}^{I_1I_2})_{ijkl}(Q_{{\bf D}}^{I_1I_2})_{ijkl} 
$$ 
and, therefore,
$$
(P_{\bf D})_{I_1I_2I_3I_4}=\frac{(Q_{{\bf D}}^{I_1I_2})_{ijkl}(Q_{{\bf D}}^{I_3I_4})_{ijkl}}
{(Q_{{\bf D}}^{IJ})_{mnsp}(Q_{{\bf D}}^{IJ})_{mnsp}}.
$$

In this way one obtains the following explicit expressions for projectors
singling out the contributions of the irreps:
\bea
\nonumber
(P_{\bf 15})_{I_1I_2I_3I_4}&=&-\frac{1}{30}C^{-}_{I_1I_2I_3I_4},\\
\nonumber
(P_{{\bf 20}})_{I_1I_2I_3I_4}&=&\frac{3}{100}\left(C^{+}_{I_1I_2I_3I_4}
-\frac{1}{6}\d_{I_1I_2}
\d_{I_3I_4}\right),\\
\nonumber
(P_{\bf 84})_{I_1I_2I_3I_4}&=&\frac{1}{504}\left(2\d_{I_1I_3}\d_{I_2I_4}
+2\d_{I_1I_4}\d_{I_2I_3}
+\frac{1}{5}\d_{I_1I_2}\d_{I_3I_4}-4C_{I_1I_3I_2I_4}
-2C^+_{I_1I_2I_3I_4}\right),\\
\nonumber
(P_{\bf 105})_{I_1I_2I_3I_4}&=&\frac{1}{1260}\left(
2\d_{I_1I_3}\d_{I_2I_4}+2\d_{I_1I_4}\d_{I_2I_3}
+\frac{1}{5}\d_{I_1I_2}\d_{I_3I_4}
+8C_{I_1I_3I_2I_4}-\frac{16}{5}C^{+}_{I_1I_2I_3I_4}\right),\\
\nonumber
(P_{\bf 175})_{I_1I_2I_3I_4}&=&\frac{1}{350}
\left(\d_{I_1I_3}\d_{I_2I_4}-
\d_{I_1I_4}\d_{I_2I_3}+ C^-_{I_1I_2I_3I_4}\right).
\eea
%%%%%%%%%%%%%%%%%%%%%%%%%%%%%%%%%%%%%%%%%%%%%%%%%%%%%%%%%%%%%%%%%%%%%%%%%%%

Defining also $P_{\bf 1}=\frac{1}{400}\d_{I_1I_2} \d_{I_3I_4}$, one can
verify that the tensors $P_{\bf 1}$,...,$P_{\bf 175}$ project an arbitrary 4-point function of
operators in the irrep {\bf 20} into the corresponding six subspaces of the direct
sum decomposition ${\bf 20}\times {\bf 20} = {\bf 1} + {\bf 20} + {\bf 84} + {\bf 105} + {\bf 15}+ 
{\bf 175}$.

The following formulae
\ba
\nonumber
&&C_{I_1I_2I_3I_4}C_{I_1I_2I_3I_4}=\frac{380}{3},~~~
C_{I_1I_2I_3I_4}C_{I_2I_1I_3I_4}=\frac{20}{3},~~~ 
C^-_{I_1I_2I_3I_4}C_{I_1I_3I_2I_4}=0,\\
\nonumber
&&C^+_{I_1I_2I_3I_4}C^+_{I_1I_2I_3I_4}=\frac{200}{3},~~~
C^+_{I_1I_2I_3I_4}C^-_{I_1I_2I_3I_4}=0,~~~
C^+_{I_1I_2I_3I_4}C_{I_1I_3I_2I_4}=\frac{20}{3}.
\ea
are helpful to find the contractions of 
the projectors with tensors describing the 
4-point function. The results for contractions 
are summarized in the Table 1. 
\begin{center}
\begin{tabular}{|c|c|c|c|c|c|}
\hline 
& & & & & \\
Tensor&$C^{+}_{I_1I_2I_3I_4}$ & $C^{-}_{I_1I_2I_3I_4}$ 
&$C_{I_1I_3I_2I_4}$ &$\d_{I_1I_3}\d_{I_2I_4}$ 
&$\d_{I_1I_4}\d_{I_2I_3}$ \\
& & & & & \\  \hline
& & & & & \\
$\frac{1}{400}\d_{I_1I_2}\d_{I_3I_4}$ & $\frac{1}{6}$ & 0 & 
$\frac{1}{60}$ & $\frac{1}{20}$ & $\frac{1}{20}$ \\
& & & & & \\ \hline
& & & & & \\
$(P_{\bf 15})_{I_1I_2I_3I_4}$ & 0 & -2 & 0 & 1 & -1 \\
& & & & & \\ \hline
& & & & & \\
$(P_{\bf 20})_{I_1I_2I_3I_4}$ & $\frac{5}{3}$ & 0 & 
$\frac{1}{6}$ & 1 & 1 \\
& & & & & \\ \hline
& & & & & \\
$(P_{\bf 84})_{I_1I_2I_3I_4}$ & 0 & 0 & 
$-\frac{1}{2}$ & 1 & 1 \\
& & & & & \\ \hline
& & & & & \\
$(P_{\bf 105})_{I_1I_2I_3I_4}$ & 0 & 0 & 1 & 1 & 1 \\
& & & & & \\ \hline
& & & & & \\
$(P_{\bf 175})_{I_1I_2I_3I_4}$ & 0 & 0 & 0 & 1 & -1 \\
& & & & & \\ \hline
\end{tabular}
\end{center}
\begin{center}
Table 1. The values of contractions of the 
projectors with tensors describing 
the structure of the 4-point function of the lowest weight CPOs.   
\end{center}

%%%%%%%%%%%%%%%%%%%%%%%%%%%%%%%%%%%%%%%%%%%%%%%%%%%%%%%%%%%%%%%%%%%%%%
\section{Appendix E. Conformal block of the conserved 2nd rank tensor}
\setcounter{equation}{0}
%%%%%%%%%%%%%%%%%%%%%%%%%%%%%%%%%%%%%%%%%%%%%%%%%%%%%%%%%%%%%%%%%%%%%
Here we sketch the derivation of the conformal block of the 
conserved second rank tensor. We do not use this result in the paper, 
however, we feel that it might be useful for subsequent studies 
of the OPE.  

We start with (\ref{OPEaa}) and suppress the 
unessential indices $\a$ and $\b$.
The 3-point function is given by the following expression
\bea
\nonumber
&&\langle O(x)O(0)T_{\mu\nu}(y) \rangle=
\frac{1}{(x^2)^{\frac{1}{2}(2\D-\D_T+2)}
(y^2)^{\frac{1}{2}(\D_T+2)}((y-x)^2)^{\frac{1}{2}(\D_T+2)}}
\biggl[\frac{\d_{\mu\nu}}{d}x^2y^2(y-x)^2 \\
\nonumber
&&-y^4(y-x)_{\mu}(y-x)_{\nu}-(x-y)^{4}y_{\mu}y_{\nu}
+y^2(y-x)^2((y-x)_{\mu}y_{\nu}+y_{\mu}(y-x)_{\nu})
\biggr],
\eea
where for simplicity we choose the 
constant $C_{OOT}$ to be equal to unity.
Compatibility of the 3-point function with 
the conservation law requires the dimension 
of the tensor to be canonical, i.e. $\D_T=d$, 
where $d=2\eta$ is a space-time dimension. 
However, in what follows we meet 
certain divergences and that is why we keep 
in some places $\eps=d-\D_T$ 
as a regularization parameter.
Substituting eq.(\ref{OPEaa}) into the 
3-point function, we get an equation 
defining the conformal block
\bea
\nonumber
&&-\frac{1}{\D_T(\D_T-2)}\left(\frac{1}{(y^2)^{\frac{1}{2}(\D_T-2)}}
e^{x\partial_y}\partial_{\mu}\partial_{\nu}
\frac{1}{(y^2)^{\frac{1}{2}(\D_T-2)}}
+\partial_{\mu}\partial_{\nu}\frac{1}{(y^2)^{\frac{1}{2}(\D_T-2)}}
e^{x\partial_y}\frac{1}{(y^2)^{\frac{1}{2}(\D_T-2)}}\right) \\
\nonumber
&&+\frac{1}{(\D_T-2)^2}\left(\partial_{\mu}
\frac{1}{(y^2)^{\frac{1}{2}(\D_T-2)}}
e^{x\partial_y}\partial_{\nu}\frac{1}{(y^2)^{\frac{1}{2}(\D_T-2)}} 
+\partial_{\nu}\frac{1}{(y^2)^{\frac{1}{2}(\D_T-2)}}
e^{x\partial_y}\partial_{\mu}\frac{1}
{(y^2)^{\frac{1}{2}(\D_T-2)}}\right)\\
\nonumber
&&-\frac{d-\D_T}{d\D_T}\d_{\mu\nu}
\left(\frac{1}{(y^2)^{\frac{1}{2}(\D_T-2)}}
e^{x\partial_y}\frac{1}{(y^2)^{\frac{1}{2}\D_T}}
+\frac{1}{(y^2)^{\frac{1}{2}\D_T}}
e^{x\partial_y}\frac{1}{(y^2)^{\frac{1}{2}(\D_T-2)}}\right)\\
\nonumber 
&&+\frac{2\d_{\mu\nu}}{d(\D_T-2)^2}\partial_{\lambda}
\frac{1}{(y^2)^{\frac{1}{2}(\D_T-2)}}
e^{x\partial_y}\partial^{\lambda}
\frac{1}{(y^2)^{\frac{1}{2}(\D_T-2)}}
=
\sum_{k=2}^{\infty} 
\frac{1}{k!}\D^k_{\rho\lambda}(x,\partial_y)\langle T_{\rho\lambda}(0)T_{\mu\nu}(y)\rangle , 
\eea
where we have introduced the following representation
$$
C_{\mu\nu}(x,\partial_y)=\sum_{k=2}^{\infty} \frac{1}{k!}\D^k_{\rho\lambda}(x,\partial_y).
$$
Using the series representation for the 
exponentials on the l.h.s. of the equation 
defining the conformal block, one then obtains an equation for $\D^k_{\rho\lambda}(x,\partial_y)$.

The 2-point function of the second rank tensor 
can be written in the form \cite{OP,EO} 
\bea
\nonumber
\langle T_{\rho\lambda}(0)T_{\mu\nu}(y)\rangle
&=&\frac{1}{(\D_T-3)(\D_T-2)\D_T(\D_T+1)}{\cal E}_{\mu\nu\rho\lambda;\a\b\gamma\delta}
\partial_{\a}\partial_{\b}\partial_{\g}\partial_{\d}\frac{1}{(y^2)^{\D_T-2}},
\eea
where again for simplicity we choose the constant $C_T$ to be the unity. 
Here ${\cal E}_{\mu\nu\rho\lambda;\a\b\gamma\delta}$ 
is a tensor with the following structure  
$$
{\cal E}_{\mu\nu\rho\lambda;\a\b\gamma\delta}\partial_{\a}
\partial_{\b}\partial_{\g}\partial_{\d}
=\frac{\D_T-3}{4(\D_T-2)}\left(
\frac{1}{2}\Box^2(\d_{\mu\rho}\d_{\nu\lambda}+\d_{\mu\rho}\d_{\nu\lambda})
+\d_{\rho\lambda}(...)+\mbox{longitudinal}\right).
$$
If we suppose that the conformal block 
acting on the 2-point function is  
symmetric traceless and transversal, 
then only the first term here is of importance
and one gets 
\bea
\nonumber
\D^k_{\rho\lambda}(x,\partial_y)\langle T_{\rho\lambda}(0)T_{\mu\nu}(y)\rangle
=\frac{4(\D_T-1)(\D_T-1-\eta)(\D_T-\eta)}{(\D_T-2)\D_T(\D_T+1)}
\D^k_{\mu\nu}(x,\partial_y)\frac{1}{(y^2)^{\D_T}}.
\eea 
Now we substitute every function $\frac{1}{(y^2)^a}$ 
appearing in the equation defining 
the conformal block for its Fourier transform  
\bea
\nonumber
\frac{1}{(y^2)^{a}}=2^{2(\eta-a)}\pi^{\eta}
\frac{\G(\eta-a)}{\G(a)}\frac{1}{(2\pi)^{2\eta}}
\int \rmd p\frac{e^{-ipy}}{(p^2)^{\eta-a}}
\eea
and find\footnote{We keep $\eps$ only there where it is 
actually needed to compute the limit $\eps\to 0$.}
\bea
\nonumber
\D^k_{\mu\nu}(x,-ip)&=&-\frac{2^4}{\pi^{\eta}}
\frac{\G(2\eta)(2\eta+1)}{\G^2(\eta-1)\G(2-\eta)}(p^2)^{-\eta}
\int \rmd q\frac{(p_{\mu}-q_{\mu})(p_{\nu}-q_{\mu})}{((p-q)^2q^2)^{1+\eps/2}}(-ixq)^k. 
\eea 
Since the conformal block is applied to the traceless transversal 
operator (2-point function) in the last expression 
we have omitted all trace and longitudinal terms 
proportional to $\d_{\mu\nu}$ and  
to $p_{\mu}$ respectively. The equation  can be then brought to the form
\bea
\label{tb1}
\D^k_{\mu\nu}(x,-ip)&=&-\frac{4}{\pi^\eta}
\frac{\G(2\eta)(2\eta+1)}{\G^2(\eta-1)\G(2-\eta)}
\frac{(p^2)^{-\eta}}{\frac{\eps}{2}\left( \frac{\eps}{2}-1\right)}
\partial_{\mu}\partial_{\nu}I_k\left(\frac{\eps}{2}-1;\frac{\eps}{2}+1\right),
\eea 
again modulo unessential trace and longitudinal terms.
Here we introduced the following integral
\bea
\nonumber
I_k(\a_1;\a_2)=\int \rmd q
\frac{(-ixq)^k}{((p-q)^2)^{\a_1}(q^2)^{\a_2}}
\eea
that is explicitly evaluated to give
\bea
\nonumber
I_k(\a_1;  \a_2)&=&\frac{\pi^{\eta}}{\G(\a_1)\G(\a_2)}
\sum_{m=0}^{[k/2]} {k \choose 2m} 
\frac{(2m)!}{m!}(-ixp)^{k-2m}\left(-\frac{1}{4}x^2p^2\right)^m 
(p^2)^{\eta-\a_1-\a_2} \\
\nonumber
&\times&\frac{\G(k-m+\eta-\a_2)\G(m+\eta-\a_1)}{\G(k+2\eta-\a_1-\a_2)}
\G(\a_1+\a_2-m-\eta).
\eea  
Again neglecting the trace and longitudinal 
contributions, we evaluate the limit 
$\eps\to 0$ and normalize the resulting 
expression such that the first nontrivial term 
$\D_{\mu\nu}^2$ starts as $\D_{\mu\nu}^2(x,-ip) = 
2x_{\mu}x_{\nu} + \ldots$. In this way 
we find the following expression 
\bea
\nonumber
\D^k_{\mu\nu}(x,\partial_y)&=&x_{\mu}x_{\nu}\frac{\G(2\eta+2)}{\G(\eta+1)}\\
\nonumber
&\times&\sum_{m=0}^{[k/2]-1}
\frac{k!}{(k-2m-2)!m!}\frac{\G(k-m+\eta-1)}{\G(k+2\eta)}
(x\partial_y)^{k-2m-2}
\left(-\frac{1}{4}x^2\D_y\right)^m .
\eea
Finally, performing the summation in $k$ 
we recover the expression $C_{\mu\nu}(x,\partial_y)$
for the conformal block of the conserved 
second rank tensor given in Appendix A.


\newpage
\begin{thebibliography}{99}
{\small
\bibitem{M} J. Maldacena, 
``The large $N$ limit of superconformal field theories and
supergravity'', Adv. Theor. Math. Phys. {\bf 2} (1998) 231.
\bibitem{GKP} G.G. Gubser, I.R. Klebanov and A.M. Polyakov, 
``Gauge theory correlators from
noncritical string theory'', 
Phys.Lett. {\bf B428} (1998) 105, hep-th/9802109.
\bibitem{W} E. Witten, ``Anti-de Sitter space and holography'', 
Adv.Theor.Math.Phys.  {\bf 2} (1998) 253, hep-th/9802150.
%%%%%%%%%%%%%%%%%%%%%%%%%%%%%%%%%%%%%%%%%%%%%%%%%%%%%%%%%%%%%%%%
\bibitem{AF1} G. Arutyunov and S. Frolov, 
``On the origin of the supergravity boundary
terms in the AdS/CFT correspondence'',
Nucl.Phys. {\bf B544} (1999) 576, hep-th/9806216.
%%%%%%%%%%%%%%%%%%%%%%%%%%%%%%%%%%%%%%%%%%%%%%%%%%%%%%%%%%%%%%%
\bibitem{S} J.H. Schwarz, ``Covariant field equations of chiral $N=2$
$D=10$ supergravity'', Nucl.Phys. {\bf B226} (1983) 269.
\bibitem{SW} J.H. Schwarz and P.C.West, ``Symmetries and
  transformations of chiral   
$N=2$ $D=10$ supergravity'', Phys.Lett. {\bf B126} (1983) 301.
\bibitem{HW} P.S. Howe and P.C. West, ``The complete  
$N=2$ $D=10$ supergravity'', Nucl.Phys. {\bf B238} (1984) 181.
%%%%%%%%%%%%%%%%%%%%%%%%%%%%%%%%%%%%%%%%%%%%%%%%%%%%%%%%%%%%%%%%%%
\bibitem{AF3} G. Arutyunov and S. Frolov, ``Quadratic action for 
type IIB supergravity on $AdS_5\times S^5$'', JHEP {\bf 08} (1999) 024,
 hep-th/9811106.
\bibitem{LMRS} S. Lee, S. Minwalla, M. Rangamani, N. Seiberg,
``Three-point functions of chiral operators in D=4'', ${\cal N}=4$ SYM
at Large N 
Adv. Theor. Math. Phys. {\bf 2} (1998) 697, hep-th/9806074.
\bibitem{AF5} G. Arutyunov and S. Frolov, ``Some cubic couplings
in type IIB supergravity on $AdS_5\times S^5$ and three-point functions
in SYM$_4$ at large $N$'', Phys.Rev. {\bf D61} (2000) 064009, hep-th/9907085.
\bibitem{Lee} S. Lee, ``AdS(5)/CFT(4) four-point functions of chiral
primary operators: Cubic vertices'', Nucl.Phys. {\bf B563}  (1999)
349, hep-th/9907108.  
\bibitem{FMMR} ``D. Freedman, S. Mathur, A. Matusis and L. Rastelli,
Correlation functions in the CFT(d)/AdS(d+1) correspondence'',
Nucl. Phys. {\bf B546} (1999) 96, hep-th/9804058.
%%%%%%%%%%%%%%%%%%%%%%%%%%%%%%%%%%%%%%%%%%%%%%%%%%%%%%%%%%%
\bibitem{LT1} H. Liu and A.A. Tseytlin, 
``On four-point functions in the CFT/AdS Correspondence'',
Phys.Rev. {\bf D59} (1999) 086002, 
hep-th/9807097.
\bibitem{FMMR1} D. Freedman, S.D. Mathur, A. Matusis and L. Rastelli, 
``Comments on 4-point functions in the CFT/AdS correspondence'', 
Phys.Lett. {\bf B452} (1999) 61,
hep-th/9808006.
\bibitem{CS} G. Chalmers and K. Schalm,
``The large $N_c$ limit of four-point functions in 
$N=4$ super Yang-Mills theory from
Anti-de Sitter supergravity'', hep-th/9810051.
\bibitem{HF} E. D'Hoker and D. Freedman, 
``Gauge boson exchange in $AdS_{d+1}$'', Nucl.Phys. {\bf B544} (1999) 612,
hep-th/9809179. 
\bibitem{BG} J.H. Brodie and M. Gutperle, "String corrections
to 4-point functions in the AdS/CFT correspondence",
Phys.Lett. {\bf B445} (1999) 296, hep-th/9809067.
\bibitem{L} H. Liu, ``Scattering in Anti-de Sitter space and
operator product expansion'', hep-th/9811152.
\bibitem{HF1} E. D'Hoker and D. Freedman, ``General Scalar Exchange in
  AdS$_d+1$'', 
Nucl.Phys. {\bf B550} (1999) 612, hep-th/9811257.
\bibitem{HFMMR} E. D'Hoker, D. Freedman, S. Mathur, A. Matusis and
  L. Rastelli, 
``Graviton exchange and complete 4-point functions in 
the AdS/CFT correspondence'', hep-th/9903196.
\bibitem{HFR} E. D'Hoker, D. Freedman and L. Rastelli,
``AdS/CFT 4-point functions: How to succeed at $z$-integrals
without really trying'', hep-th/9905049. 
\bibitem{Sa} Sanjay, ``On direct and crossed channel asymptotics of 
four-point functions in AdS/CFT correspondence'', Mod.Phys.Lett. {\bf
  A14}  (1999) 1413,
hep-th/9906099.
\bibitem{HMMR}  E. D'Hoker, S.D. Mathur, A. Matusis and L. Rastelli, 
``The Operator Product Expansion of ${\cal N}=4$ SYM and the 4-point 
Functions of Supergravity'', hep-th/9911222.
\bibitem{HPR1} L. Hoffmann, A.C. Petkou and W. R\"uhl, ``A note on the
  analyticity 
of AdS scalar exchange graphs in the crossed channel'', to appear in
Phys. Lett. {\bf B}, hep-th/0002025.
\bibitem{H} C.P. Herzog, `` OPEs and 4-point functions in AdS/CFT
  correspondence'',   
hep-th/0002039.
\bibitem{HPR2} L. Hoffmann, A.C. Petkou and W. R\"uhl, ``Aspects of the
  conformal  
Operator Product Expansion in AdS/CFT correspondence'', hep-th/0002154.
%%%%%%%%%%%%%%%%%%%%%%%%%%%%%%%%%%%%%%%%%%%%%%%%%%%%%%%%%%%%%%%%%%%%%
\bibitem{AF6} G. Arutyunov and S. Frolov, ``Scalar Quartic Couplings
in Type IIB Supergravity on $AdS_5\times S^5$'', 
to appear in Nucl. Phys. {\bf B}., hep-th/9912210.
\bibitem{AF7} G. Arutyunov and S. Frolov,
``Four-point Functions of Lowest Weight CPOs in $\N =4$ SYM$_4$ in 
Supergravity Approximation'', to appear in Phys. Rev. {\bf D}., 
hep-th/0002170.
\bibitem{Sym} K. Symanzik, ``Small-distance-behaviour analysis 
and Wilson expansions'', Commun. Math. Phys. {\bf 23} (1971) 49.
\bibitem{LR} K. Lang and W. R\"uhl, ``The critical  $O(N)$
$\sigma$-model at dimension $2<d<4$: a list of quasi-primary fields'',
Nucl. Phys. {\bf B402} (1993) 573; ``The critical $O(N)$
$\sigma$-model at dimension $2<d<4$ order $1/N^2$: Operator product
expansions and renormalization'', 
Nucl. Phys. {\bf B377} (1992) 371.
\bibitem{P} A. C. Petkou, ``Conserved currents, consistency relations
and operator product expansions in the conformally invariant $O(N)$
vector model'', Ann. Phys. {\bf 249} (1996) 180, hep-th/9410093. 
\bibitem{AFer} L. Andrianopoli and S. Ferrara, ``On short and long SU(2,2/4)
in the AdS/CFT correspondence'', hep-th/9812067. 
\bibitem{AFer2} S. Ferrara and A. Zaffaroni, ``Superconformal
Field Theories, Multiplet Shortening and the 
AdS$_5$/SCFT$_4$ Correspondence'', hep-th/9908163.
%%%%%%%%%%%%%%%%%%%%%%%%%%%%%%%%%%%%%%%%%%%%%%%%%%%%%%%%%%%%%%%%
\bibitem{EHSSW} B. Eden, P.S. Howe, C. Schubert, E. Sokatchev and
P.C. West, ``Simplifications of four-point functions in $\N=4$
supersymmetric Yang-Mills theory at two loops'', hep-th/9906051.
\bibitem{BKRS1} M. Bianchi, S. Kovacs, G. Rossi and Y.S. Stanev, 
``On logarithmic behaviour in ${\cal N}=4$ SYM theory'',
hep-th/9906188.
\bibitem{ESS} B. Eden, C. Schubert and E. Sokatchev, ``Three-loop
four-point correlator in $\N=4$ SYM'', hep-th/0003096.
\bibitem{BKRS2} M. Bianchi, S. Kovacs, G. Rossi and Y.S. Stanev, 
``Anomalous dimensions in $\N =4$ SYM at order $g^4$'',
hep-th/0003203.
%%%%%%%%%%%%%%%%%%%%%%%%%%%%%%%%%%%%%%%%%%%%%%%%%%%%%%%%%%%%%%%%%%%%%%%
\bibitem{An} D. Anselmi, ``The $N=4$ quantum conformal algebra'',
Nucl. Phys. {\bf B541} (1999) 369, hep-th/9809195.
\bibitem{An1} D. Anselmi, ``Quantum conformal algebras and closed
  conformal field theory'', 
Nucl. Phys. {\bf B554} (1999) 415, hep-th/9811149.
%%%%%%%%%%%%%%%%%%%%%%%%%%%%%%%%%%%%%%%%%%%%%%%%%%%%%%%%%%%%%%%%%%%%%%%
\bibitem{OP} H. Osborn and A. Petkou, 
``Implications of conformal invariance in field theories for general
dimensions'', 
Ann. Phys. (N.Y.) {\bf 231} (1994) 311, hep-th/9307010.
\bibitem{MPet}
R. Manvelyan and A. C. Petkou, ``$R$-current anomalies in the
$(2,0)$ tensor multiplet in $d=6$ and AdS/CFT correspondence'',
hep-th/000310, to appear in Phys. Lett. {\bf B}.
\bibitem{EO} J. Erdmenger and H. Osborn, 
``Conserved currents and the energy momentum tensor in conformally invariant
theories for general dimensions'',
Nucl. Phys. {\bf B483} (1997) 431, hep-th/9605009.
\bibitem{LT} H. Liu and A. A. Tseytlin, 
``D=4 Super-Yang-Mills, D=5 gauge supergravity and
D=4 conformal supergravity'',
Nucl.Phys. {\bf B553} (1998) 88, hep-th/9804083.
\bibitem{Sk} W. Skiba, ``Correlators of Short Multi-Trace Operators in
$N=4$ Supersymmetric Yang-Mills'', Phys. Rev. {\bf D60} (1999) 
105038, hep-th/9907088.
%%%%%%%%%%%%%%%%%%%%%%%%%%%%%%%%%%%%%%%%%%%%%%%%%%%%%%%%%%%%%%%%%%%%%%%%%%%%%%
\bibitem{Mack}
G. Mack, ``Group theoretical approach to conformal invariant
quantum field theory'', in: "{\it Renormalization and invariance in
quantum field theory}", ed. Caianiello, Plenum Press, New York
(1974); V. K. Dobrev, V. B. Petkova, S. G. Petrova and I. T. Todorov,
``Dynamical derivation of vacuum operator product
expansion in Euclidean conformal quantum field theory'', 
Phys. Rev. {\bf D13} (1976) 887. 
\bibitem{Fradkin}
E. S. Fradkin and M. Ya. Palchik, ``New developments in
$d$-dimensional conformal quantum field theory'', Phys. Rep. {\bf 300}
(1998) 1. 
\bibitem{FGGP} S. Ferrara, R. Gatto and A. F. Grillo, ``Conformal
invariance on the light-cone and canonical dimensions'', 
Nucl. Phys. {\bf B34} (1971) 349; ``Tensor representations of
conformal covariant 
operator product expansions'', Ann. Phys. {\bf 76} (1973) 161.\\
S. Ferrara, R. Gatto, A. F. Grillo and G. Parisi, 
``Covariant expression of the conformal four-point function'', 
Nucl. Phys. {\bf B49} (1972) 77.
}   
\end{thebibliography}
\end{document}